\documentclass[draft,monochrome]{agujournal2019}
\usepackage{url} 
\usepackage[inline]{trackchanges} 
\usepackage{soul}
\usepackage{colortbl}
\usepackage{tcolorbox}
\usepackage{{amsmath,amsfonts,amssymb}}
\DeclareMathOperator{\pt}{\partial}

\usepackage{lastpage}  
\usepackage[explicit]{titlesec}

\drafttrue

\journalname{} 
\graphicspath{{.}{pics}}

\begin{document}

%
%


\title{Learning Physically Interpretable {\color{red}Atmospheric} Models from Data with WSINDy}

%
%




\authors{Seth Minor\affil{1}, Daniel A.~Messenger\affil{2}, Vanja Dukic\affil{1}$^{*}$, and David M.~Bortz\affil{1}\thanks{Joint senior authors}}

\affiliation{1}{Department of Applied Mathematics, University of Colorado, Boulder, CO 80309-0526 USA}
\affiliation{2}{Theoretical Division, Los Alamos National Laboratory, Los Alamos, NM 87545 USA}




\correspondingauthor{Seth Minor}{seth.minor@colorado.edu}
\correspondingauthor{David Bortz}{david.bortz@colorado.edu}



\begin{keypoints}
\item We review the WSINDy algorithm and illustrate {\color{blue} how} it can be used to {\color{red}discover} interpretable mathematical models from noisy {\color{red}atmospheric} data.
\item We {\color{blue}demonstrate the discovery of governing equations for} three simulated {\color{blue} datasets of geophysical interest}.

\item {\color{blue}Additionally, we} apply WSINDy to global-scale assimilated data sourced from the European Center for Medium-Range Weather Forecasts.
\end{keypoints}

%
%

%
%


\begin{abstract}
The multiscale and turbulent nature of Earth's atmosphere has historically rendered accurate weather modeling a hard problem. Recently, there has been an explosion of interest surrounding data-driven approaches to weather modeling, which in many cases show improved forecasting accuracy and computational efficiency when compared to traditional methods. However, many of the current data-driven approaches employ highly parameterized neural networks, often resulting in uninterpretable models and limited gains in scientific understanding. In this work, we address the interpretability problem by explicitly discovering partial differential equations governing {\color{red}atmospheric} phenomena, identifying symbolic mathematical models with direct physical interpretations. The purpose of this paper is to demonstrate that, in particular, the Weak form Sparse Identification of Nonlinear Dynamics (WSINDy) algorithm can learn effective {\color{red}atmospheric} models from both simulated and assimilated data. Our approach adapts the standard WSINDy algorithm to work with high-dimensional fluid data of arbitrary spatial dimension. 
\end{abstract}

\section*{Plain Language Summary}
The Weak form Sparse Identification of Nonlinear Dynamics (WSINDy) algorithm is a recently-developed computational method which can be used to learn physically interpretable models directly from noisy data. In this work, we demonstrate how WSINDy can be used to discover explicit mathematical models for complicated {\color{red}atmospheric} {\color{blue}phenomena}.


\newpage

%
%

%


%
%
%
%

\section{Introduction and Previous Work}

Since its modern inception in the pioneering computational work of Charney, Fj{\"o}rtoft, and Von Neumann \cite{CharneyFjortoftNeumann1950Tellus}, numerical weather prediction (NWP) has proven to present formidable mathematical challenges. In particular, many dynamic models of weather phenomena exhibit multiscale and turbulent solutions which have been known since the seminal work of \citeA{Lorenz1963JAtmosSci}
to lead to a sensitive dependence on initial conditions. As a consequence, any errors present in a set of initial observations grow exponentially in time under these models, bounding the predictive power of most numerical weather forecasts to medium-range time scales ($\leq$ 14 days). This chaotic behavior is exacerbated by the reality that simulations of the relevant physics can only capture a finite range of scales, so that the physical influence of unresolved scales is either ignored or approximated by subgrid closure models.

In recent years, there has been an explosion of interest surrounding data-driven approaches to {\color{red}atmospheric} modeling{\color{red};} see, e.g., \citeA{RaspHoyerMeroseEtAl2024JAdvModelEarthSyst} and \citeA{KarlbauerMaddixAnsariEtAl2024arXiv240714129} for a discussion of recent benchmarks. In contrast to traditional NWP, which relies on numerical simulations of physics-based weather models, these novel data-driven approaches learn effective weather models directly from empirical data. A common theme of recent work in this area is the use of highly-parameterized neural networks trained to predict future weather conditions using one of two modi operandi: (1) learn an effective model from the empirical data alone (without reference to external knowledge of the physics), or (2) incorporate physical knowledge to learn a model in a \textit{hybrid} fashion (e.g., penalizing potential models that violate known physics). State-of-the-art examples in each of these two categories are, respectively, GraphCast \cite{LamSanchez-GonzalezWillsonEtAl2023Science}, which uses graph neural networks to predict and relate weather dynamics on a range of length scales, and NeuralGCM \cite{KochkovYuvalLangmoreEtAl2024Naturea}, which uses a hybrid neural network architecture to represent parameterized physical processes included within an explicit base model. While such models achieve both (1) significant computational speedups, often by orders of magnitude, and (2) improved accuracy in forecasting over traditional methods \cite{RaspHoyerMeroseEtAl2024JAdvModelEarthSyst}, the large number of parameters renders such models almost completely uninterpretable. For example, GraphCast has roughly 36.7 million parameters \cite{LamSanchez-GonzalezWillsonEtAl2023Science}.

\newpage

A separate thread of research seeks to discover data-driven {\color{red}geophysical} models in a physically interpretable symbolic form, such as the effective governing partial differential equations (PDEs). Some recent approaches, e.g., \citeA{ZannaBolton2020GeophysicalResearchLetters}, use physical data to learn explicit subgrid closure models, which {\color{blue}can} then {\color{blue}be} appended onto {\color{red}idealized} PDEs to improve their accuracy when modeling real-world data. However, physical data can also be used to estimate the equations of motion in their entirety. A popular framework for learning PDEs from data is sparse dictionary learning as used in, e.g., the Sparse Identification of Nonlinear Dynamics (SINDy) algorithm \cite{BruntonProctorKutz2016ProcNatlAcadSci}. SINDy attempts to fit functions from a library of candidate terms to an evolution operator, while also using a regularized loss function to enforce a parsimonious solution with a relatively small number of terms. Unfortunately, the na\"ive SINDy algorithm is not robust to observational noise, potentially limiting the value of the approach in {\color{red}empirical} contexts. However, recent advancements in data-driven model discovery, such as the development of Galerkin methods like the Weak SINDy (WSINDy) algorithm of \citeA{MessengerBortz2021JComputPhys} (see also, e.g., \citeA{ReinboldKageorgeSchatzEtAl2021NatCommun} and \citeA{GurevichGoldenReinboldEtAl2024JFluidMech}), have substantially increased robustness to noise by representing and in turn learning the relevant dynamics in their weak form. In this formulation, the data are integrated against localized test functions which implicitly allow for both the extraction of signal-dominated modes and the imposition of a set of particular length and time scales.

The purpose of this paper is to demonstrate that WSINDy is a powerful tool for interpretable, data-driven geophysics. Herein, we {\color{blue}specifically} adapt WSINDy to the task of {\color{red}weather and climate} modeling, illustrating the discovery of effective PDE models from both simulated and assimilated {\color{red}atmospheric} data spanning several common meteorological regimes. We organize the paper as follows. In Section~\ref{background}, we introduce the notational conventions used throughout (\S\ref{notation}) before reviewing select background material related to mathematical weather modeling (\S\ref{governing equations}), our implementation of the WSINDy algorithm (\S\ref{wsindy}), and the performance metrics (\S\ref{metrics}) used to assess the quality of our results. We then present and discuss our model discovery results in Section~\ref{results}, detailing our choice of hyperparameters (\S\ref{implementation_details}) used to obtain results on simulated (\S\ref{simulated}) and assimilated (\S\ref{assimilated}) datasets {\color{blue}and, in turn, commenting on the resulting residual error (\S\ref{residuals}) and forecasting capacity (\S\ref{forecasting}).} Finally, in Section~\ref{conclusion}, we conclude with a brief summary of the paper and some reflections on natural extensions of this work. Supplemental information about the datasets and numerical methods used to produce these results is given in the appendix.

\section{Methods and Background Material}\label{background}
Here, we provide a brief overview of the mathematical content we primarily draw upon in applying weak form model discovery to {\color{red}atmospheric} data. For a review of PDE models {\color{red}traditionally used in weather contexts}, interested readers are directed towards the review given by \citeA{White2003EncyclopediaofAtmosphericSciences}. For a more complete discussion of weak form model discovery applied to spatiotemporal systems, readers are directed towards the recent SIAM News article by \citeA{MessengerTranDukicEtAl2024SIAMNews}, its companion review article \cite{MessengerTranDukicEtAl2024arXiv240906751}, a more technical book chapter \cite{BortzMessengerTran2024NumericalAnalysisMeetsMachineLearning}, and the original WSINDy for PDEs paper \cite{MessengerBortz2021JComputPhys}.

\subsection{Notation and Conventions}\label{notation}
In this paper, we consider $(n+1)$-dimensional dynamics on bounded spatiotemporal domains $(\boldsymbol{x}, t) \in \mathcal{X} \times [0, T]$, where $\mathcal{X} \subset \mathbb{R}^n$. When referencing planetary scale {\color{red}atmospheric} data, we use a geographic coordinate system $\boldsymbol{x} = (\varphi, \theta, r) {\color{blue}:=} \varphi\boldsymbol{\hat{\varphi}} + \theta\boldsymbol{\hat{\theta}} + r\boldsymbol{\hat{r}}$ in which $\varphi \in [0, 2\pi)$ is the longitude, $\theta \in [-\frac{\pi}{2}, \frac{\pi}{2}]$ is the latitude, and $r \geq 0$ is the altitude. For geographic coordinates, we use $u, \, v$, and $w$ to denote the zonal ({\color{blue}west-east}), meridional (north-south), and radial components of the wind velocity {\color{red}vector} $\boldsymbol{v}=(u,v,w)$ (the ERA5 dataset of \S\ref{assimilated} uses a pressure-based vertical coordinate $\eta(p)$ which satisfies $\dot{\eta}|_p \propto -w|_p$).  We denote the horizontal surface velocity as $\boldsymbol{u} = (u,v) {\color{blue}:= u\boldsymbol{\hat{\varphi}} + v\boldsymbol{\hat{\theta}}}$ and the {\color{blue}relative} vorticity as {\color{blue}$\zeta = |\!|\nabla \times \boldsymbol{u}|\!|_2$}. When necessary, we approximate the rotation rate of Earth as $\Omega = |\!|\boldsymbol{\Omega}|\!|_2 \approx 7.29 \cdot 10^{-5}$ (rads/$s$), where $\boldsymbol{\Omega}$ is the planetary angular velocity vector. The Coriolis force induced by a planetary reference frame is $\boldsymbol{f} = (-fv, fu)$, where $f := 2\Omega\sin(\theta)$. On the surface of a sphere of radius $r=a$, the presence of metric terms in the local gradient operator $\nabla = a^{-1}(\sec(\theta)\pt_{\varphi}, \ \pt_{\theta})$ entail that the advection $\mathfrak{A}(\rho)$ of a scalar field $\rho = \rho(\varphi, \theta)$ takes the form \begin{align}\label{eq:advection_operator}
    \mathfrak{A}(\rho) := \left(\boldsymbol{u} \cdot \nabla\right)\rho
    = \frac{1}{a\cos(\theta)} \Big[u\pt_{\varphi} \, + \, \cos(\theta)v\pt_{\theta}\Big] \rho,
\end{align} while the local horizontal divergence operator $\mathfrak{D}(\rho)$ is instead given by \begin{align}\label{eq:divergence_operator}
    \mathfrak{D}(\rho) := \nabla \cdot \left(\rho\boldsymbol{u}\right)
    = \frac{1}{a\cos(\theta)} \Big[\!\pt_{\varphi}(\rho u) + \pt_{\theta}\!\big(\!\cos(\theta)\rho v \big)\Big].
\end{align} For a derivation of these operators, we specifically refer the reader to the discussion leading up to eqs.~(20) and (31) in \cite{White2003EncyclopediaofAtmosphericSciences}. {\color{blue}Note that for vector fields such as $\boldsymbol{u}$, one instead computes $(\boldsymbol{u} \cdot \nabla)\boldsymbol{u} = \mathbf{J}^T\boldsymbol{u}$, where $\mathbf{J}$ denotes the Jacobian matrix of $\boldsymbol{u}$.}

\subsection{Governing Equations}\label{governing equations}
Earth's weather is predominantly influenced by the dynamics of wind in its atmosphere, which in turn evolves according to the compressible Navier-Stokes momentum equations. Posed in a rotating {\color{blue}geographic} reference frame, an idealized governing equation can be written in the form \begin{align}\label{eq:nav_stokes}
    \boldsymbol{v}_t +(\boldsymbol{v} \cdot \nabla) \boldsymbol{v}
    =
    \boldsymbol{f}
    - \nabla\left(p/\rho\right) + \nabla\cdot\boldsymbol{\tau}
    - \nabla\Phi,
\end{align} where here $p$ denotes the air pressure, $\rho$ is the air density, and $\boldsymbol{\tau}$ is the deviatoric stress tensor representing the effects of viscosity. Here, the {\color{blue}apparent gravitational potential} $\Phi$ is defined as a sum of gravitational and centrifugal terms via $\Phi = {\color{blue}\Phi_{\rm{g}} + \Phi_{\rm{c}}}$, where {\color{blue}$\Phi_{\rm{g}}$ is the Newtonian gravitational potential} and $\nabla {\color{blue}\Phi_{\rm{c}}}(\boldsymbol{x}) = \boldsymbol{\Omega} \times(\boldsymbol{\Omega} \times \boldsymbol{x})$ \cite{White2003EncyclopediaofAtmosphericSciences}. Many global scale weather models, however, use strategic simplifications of eq.~(\ref{eq:nav_stokes}) for NWP, omitting, e.g., vertical terms via hydrostatic approximations of the dynamics; again, see \citeA{White2003EncyclopediaofAtmosphericSciences} for a discussion. For reference, we detail an important example of one such \textit{primitive equation} in the appendix (see eqs.~(\ref{eq:IFSmomentum_u}) and (\ref{eq:IFSmomentum_v})), which is used in the Integrated Forecast System (IFS) of the European Center for Medium-Range Weather Forecasts \cite{ECMWF2021IFSDocumentationCY47R3}.

\subsection{The Weak SINDy Algorithm}\label{wsindy}
Given a set of observations $\mathcal{U}=\{\boldsymbol{u}(\boldsymbol{x}_m, t_m)\}_{m=1}^{M}$ of the state $\boldsymbol{u} = [u_1, \, \dots, \, u_d]$ of a spatiotemporal system, sparse dictionary learning methods for data-driven PDE discovery attempt to equate an \textit{evolution operator} $\mathcal{D}^0 u_l$ with a closed form expression consisting of functions taken from a \textit{library} $\boldsymbol{\Theta}(\mathcal{U})$ of candidate terms, \begin{align*}
    \boldsymbol{\Theta}(\mathcal{U})
    =
    \big\{ \mathcal{D}^if_j(\boldsymbol{u}) \big\}_{i,j=1}^{I,J},
\end{align*} which is evaluated over each observation $\boldsymbol{u}(\boldsymbol{x}_m ,t_m) \in \mathcal{U}$. Here, each $\mathcal{D}^i$ denotes one of $I$ distinct differential operators while each $f_j$ represents one of $J$ distinct scalar-valued functions of $\boldsymbol{u}$. {\color{blue}While one most commonly poses temporal evolution operators such as $\mathcal{D}^0 = \pt_t$ or $\mathcal{D}^0 = \pt_t^2$, we note that this is by no means required. For example, one might instead consider a material derivative $\mathcal{D}^0 = \pt_t + \mathfrak{A}$ or a spatial derivative of the form $\mathcal{D}^0 = \sum_{i=1}^{n}a_i\pt_{x_i}$. Unless stated otherwise, we will set the evolution operator $\mathcal{D}^0$ to $\pt_t$ to simplify the exposition.}

In the SINDy algorithm of \citeA{BruntonProctorKutz2016ProcNatlAcadSci}, the model discovery problem is recast as a regression problem posed over a sparse vector of coefficients $\mathbf{w} = [w_1, \dots, w_{IJ}]^T$ which weight candidate terms in the library. Although SINDy originally addressed finite-dimensional systems, subsequent work by \citeA{RudyBruntonProctorEtAl2017SciAdv} has extended it to the context of PDEs, where the central problem can be formulated as follows: \begin{align}
    \text{\tt find sparse $\mathbf{w}$ such that:}\quad
    \pt_tu_l(\boldsymbol{x}_m ,t_m) \approx
    \sum_{i=1}^{I} \sum_{j=1}^{J} w(i,j) \, \mathcal{D}^i f_j (\boldsymbol{u})(\boldsymbol{x}_m ,t_m),
    \label{eq:sindy}
\end{align} 
for each observation $m = 1, \dots, M$, where we adopt the notation $w(i,j) := w_{(i-1)J+j}$. Numerically, we restructure eq.~(\ref{eq:sindy}) as an equivalent linear system $\pt_t \mathbf{U} = \boldsymbol{\Theta}(\mathbf{U}) \mathbf{w}$ by defining $\mathbf{U} = [\mathbf{u}_1, \dots, \mathbf{u}_d] \in \mathbb{R}^{M \times d}$ as a matrix whose columns are given by vectorizing each component of the data, i.e., $\mathbf{u}_l := \texttt{vec}\{u_l{(\boldsymbol{x}_m, t_m)}\} \in \mathbb{R}^{M}$. In turn, we use a library $\boldsymbol{\Theta}(\mathbf{U}) \in \mathbb{R}^{M \times IJ}$ whose columns are given by $\Theta(i,j) = \texttt{vec}\{\mathcal{D}^if_j (\mathbf{U})\} \in \mathbb{R}^M$ with a corresponding weight {\color{blue}matrix} $\mathbf{w} = [\mathbf{w}_1, \dots, \mathbf{w}_d] \in \mathbb{R}^{IJ \times d}$. The terms in eq.~(\ref{eq:sindy}) then take the form of data matrices, schematically represented by \begin{align*}
    {\color{blue}
    \begin{bmatrix}
        \vline & & \vline
        \\
        \pt_t\mathbf{u}_1 & \dots & \pt_t\mathbf{u}_d
        \\
        \vline & & \vline
    \end{bmatrix}
    =
    \begin{bmatrix}
        \vline & & \vline \\
        \mathcal{D}^1\!f_1\!\left(\mathbf{U}\right) & \cdots & \mathcal{D}^I\!f_J\!\left(\mathbf{U}\right) \\
        \vline & & \vline
    \end{bmatrix}
    \begin{bmatrix}
        \vline & & \vline \\
        \mathbf{w}_1 & \cdots & \mathbf{w}_d \\
        \vline & & \vline
    \end{bmatrix}.
    }
\end{align*}

The optimal (sparse) {\color{blue}matrix} of coefficients $\mathbf{w}^{\star}$ is found by minimizing a regularized loss function $\mathcal{L}$, {\color{blue} leading to an optimization problem of the form}
\begin{align}\label{eq:sindyLoss}
    \mathbf{w}^{\star} = \text{arg}\!\min_{\!\!\!\!\!\mathbf{w}} \ \mathcal{L}\left(\mathbf{w}; \mathbf{U}_t, \boldsymbol{\Theta}\right),
    \quad \text{where} \quad
    \mathcal{L}\left(\mathbf{x}; \mathbf{b},\mathbf{A}\right) := \lvert\!\lvert \mathbf{b} - \mathbf{A}\mathbf{x} \rvert\!\rvert_2^2 + {\color{red}\mu}\lvert\!\lvert \mathbf{x} \rvert\!\rvert_0.
\end{align} The regularization term  ${\color{red}\mu}\lvert\!\lvert \mathbf{w} \rvert\!\rvert_{0}$ in the loss function promotes the selection of a sparse model by penalizing models with a large number of terms, where $\lvert\!\lvert \, \cdot \, \rvert\!\rvert_{\text{0}}$ denotes the $\ell_0$ \lq\lq{norm}.'' In practice, this is achieved by using iterative thresholding optimization schemes, which progressively restrict the number of columns of $\boldsymbol{\Theta}(\mathbf{U})$ available to the model; see \citeA{BruntonProctorKutz2016ProcNatlAcadSci} and \citeA{MessengerBortz2021JComputPhys}.

\begin{figure}
    \centering
    \fbox{\includegraphics[width=0.68\linewidth]{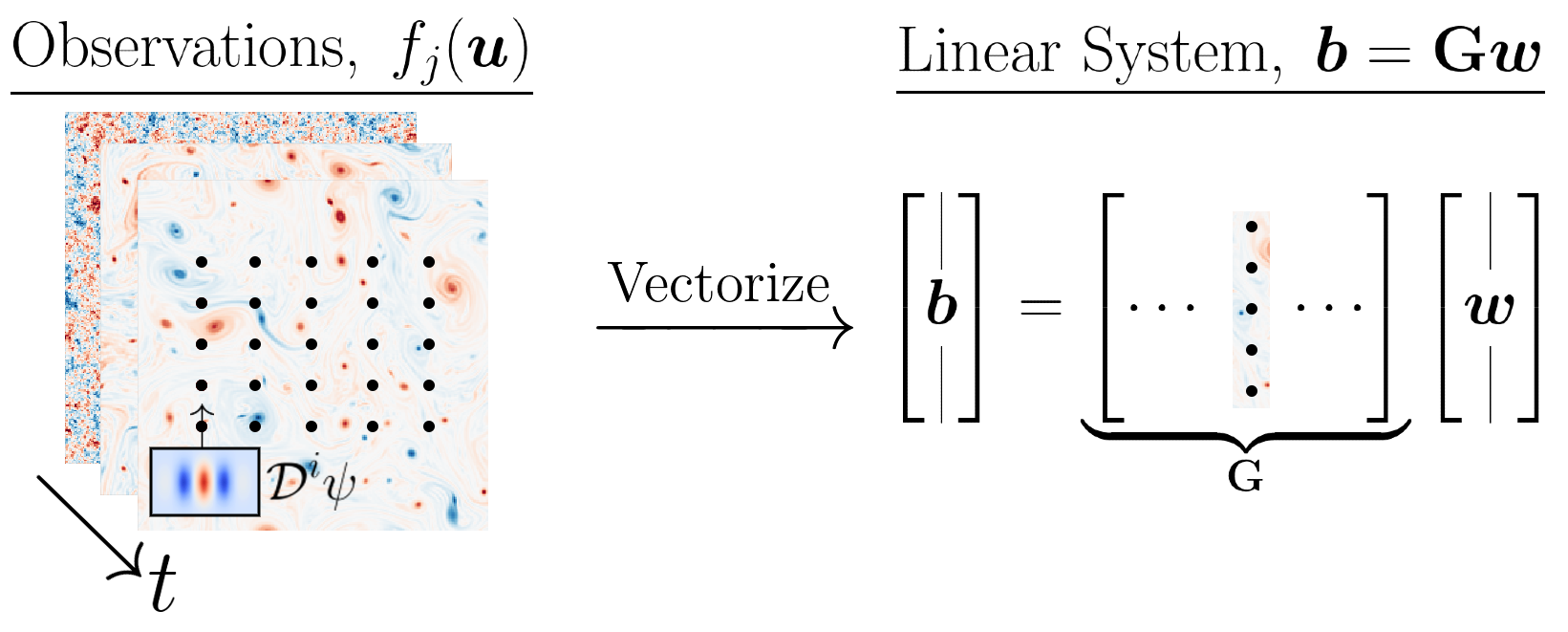}}
    \caption{A schematic illustrating the construction of the WSINDy linear system of eq.~(\ref{eq:wsindy}). Each column of the weak library $\mathbf{G}$ represents a strided convolution between a discretized test function derivative $\mathcal{D}^i\psi$ and a particular function of the data, $f_j(\boldsymbol{u})$. The result is then sub-sampled over a set of query points $\{(\boldsymbol{x}_k, t_k)\}$ (black dots) and vectorized. Here, the data are snapshots of scalar vorticity $\zeta$ from the numerical simulation of equivalent barotropic turbulence of \S\ref{barotropic}, made using PyQG \cite{AbernatheyRochanotesRossEtAl2022} code based upon \cite{Mcwilliams1984JFluidMech}.}
    \label{fig:wsindy_schematic}
\end{figure}

In the weak formulation of SINDy \cite{MessengerBortz2021JComputPhys}, the linear system in eq.~(\ref{eq:sindy}) is integrated against a collection $\{\psi_k\}_{k=1}^{K}$ of translations of a symmetric, compactly supported test function {\color{blue}$\psi \in C_c^{p}(\mathcal{X} \times [0,T])$ for sufficiently large $p$}, \begin{align*}
    {\color{blue}
    \left\langle \psi_k, \, \pt_t\!u_{l} \right\rangle
    \approx
    \sum_{i=1}^{I} \sum_{j=1}^{J} w(i,j) \, \left\langle \psi_k, \, \mathcal{D}^i f_j (\boldsymbol{u})\right\rangle,
    }
\end{align*} where each $\psi_k(\boldsymbol{x},t) := \psi(\boldsymbol{x}_k - \boldsymbol{x}, t_k - t)$ is centered at a corresponding \textit{query point} $(\boldsymbol{x}_k,t_k)$ {\color{blue}and $\langle\cdot,\cdot\rangle$ denotes the $L^2$ inner product}. A key benefit of WSINDy is that pointwise derivative computations of the data can be avoided by transferring the differential operators $\mathcal{D}^i$ from the data $f_j(\boldsymbol{u})$ to the test functions $\psi_k$ by repeated integration by parts, exploiting the compact support of the test functions:\begin{align*}
    {\color{blue}
    \left\langle \pt_t\psi_k, \, u_{l} \right\rangle
    \approx
    \sum_{i=1}^{I} \sum_{j=1}^{J} w(i,j) \, \left\langle \mathcal{D}^i\psi_k, \, f_j (\boldsymbol{u})\right\rangle.
    }
\end{align*} We note that the sign convention in the argument of each $\psi_k$ conveniently eliminates the resulting alternating factors of $(-1)^{|\alpha^i|}$, {\color{blue} where $|\alpha^i|$ is the order of the $i^{\rm{th}}$ operator}. {\color{blue}This integral formulation has been shown to exhibit substantially higher-fidelity results than SINDy in the presence of noisy data; see, e.g., Table 6 in} \cite{MessengerBortz2021JComputPhys}.

In contrast to eq.~(\ref{eq:sindyLoss}), the WSINDy weights $\mathbf{w}^{\star}$ are now found by minimizing a loss function of the flavor $\mathcal{L}(\mathbf{w}; \mathbf{b}, \mathbf{G})$; here, $\mathbf{b}$ and $\mathbf{G}$ are, respectively, real-valued $K\!\times d$ and $K\!\times\!IJ$ matrices defined by \begin{align}\label{eq:wsindy}
    \begin{cases}
        \ b_{kl} = \left(\psi_t * u_{l}\right)\!(\boldsymbol{x}_k, t_k),
        \\
        G(i,j)_k = \left(\mathcal{D}^i\psi * f_j(\boldsymbol{u})\right)\!(\boldsymbol{x}_k, t_k),
    \end{cases}
\end{align} where $*$ denotes the discrete convolution operator, {\color{blue}computed} using the trapezoidal rule on a uniformly-spaced discretized grid ${\color{blue}\mathcal{X}_{\Delta} \times T_{\Delta} := } \{(\boldsymbol{x}_m, t_m)\}_{m=1}^{M}$. This discrete convolution can be efficiently computed via the {\color{blue}FFT, giving the WSINDy algorithm an asymptotic time complexity of $\mathcal{O}(N^{n+1}\log N)$ for $N$ data points along each of $(n+1)$ dimensions}, although sparse {\color{blue}computations} may perform better when the support is sufficiently small. The construction of the linear system ${\color{blue}\mathbf{b} = \mathbf{G}\mathbf{w}}$ in eq.~(\ref{eq:wsindy}) is illustrated in Figure~\ref{fig:wsindy_schematic}.

\begin{figure}
    \centering
    \includegraphics[width=\linewidth]{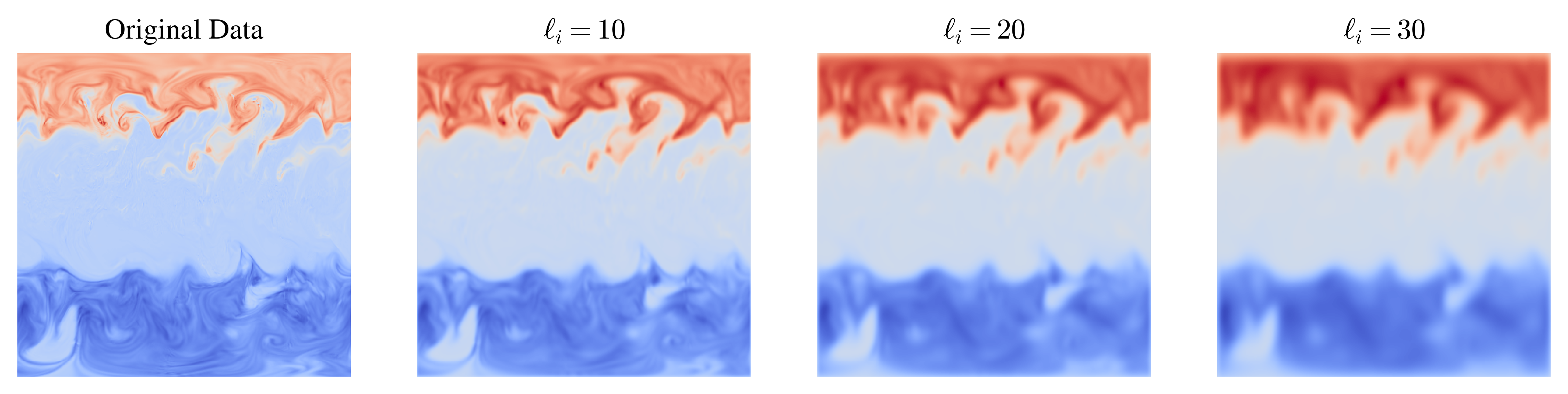}
    \caption{Illustrating the effect of increasing the test function support parameter $\ell_i$ in the convolution $(\psi * u_l)$ to select features at increasingly coarsened length scales. Here, the potential vorticity $u_l = \omega$ from the ERA5 dataset (\S\ref{era5_description}) is plotted at $t=45$ hours. This and other material in this paper contain modified Copernicus Atmosphere Monitoring Service information [2024].}
    \label{fig:changing_scales}
\end{figure}

A unique feature of integrating the data against localized test functions $\psi_k$ is that a particular set of length and time scales can be imposed upon the observed data $\mathcal{U}$ by choosing a generating test function $\psi$ with compact support given by\begin{align*}
    \text{supp}(\psi) = \big[-\ell_{x_1}\Delta{x_1}, \ \ell_{x_1}\Delta{x_1}\big] \times \cdots \times \big[-\ell_{x_n}\Delta{x_n}, \ \ell_{x_n}\Delta{x_n}\big] \times \big[-\ell_{t}\Delta{t}, \ \ell_{t}\Delta{t}\big],
\end{align*} where the $(n+1)$-tuple $\boldsymbol{\ell} = (\ell_{x_1}, \dots, \ell_{x_n}, \ell_t)$ is a tunable hyperparameter and each $\Delta_i$ represents the spacing of the discretized grid along the $i^{\rm{th}}$ axis. This allows one to discover effective models for data approximately projected onto the scales $x_i \approx \ell_{x_i}\Delta{x_i}$ in space and $t \approx \ell_t\Delta{t}$ in time, which is useful for modeling scale-dependent physical mechanisms. We note that this notion of \lq{scale}' bears a strong resemblance to the same term as used in Scale-Space Theory.

Figure~\ref{fig:changing_scales} illustrates how a progressive coarsening of the weak form data $(\psi * u_l)$ can be achieved by uniformly increasing $\ell_i$ (here, we use identical $\ell_i$ for each axis). In this instance, the resulting coarse-grained models become increasingly advection dominated. {\color{blue}Note that as the support radii $\boldsymbol{\ell} \rightarrow (0, \dots, 0)$, the WSINDy algorithm collapses to the strong formulation of the SINDy algorithm; in particular, the test functions $\psi_k$ converge to Dirac delta functions $\delta(\boldsymbol{x}_k,t_k)$ and the discretized derivatives $\mathcal{D}^i \psi(\mathcal{X}_{\Delta} \times T_{\Delta})$ converge towards familiar finite difference kernels.}

\newpage

\subsubsection{Selecting Candidate Terms}
When creating a library $\boldsymbol{\Theta}(\mathbf{U})$ of candidate terms $\{\mathcal{D}^if_j(\mathbf{U})\}$, one can reduce the numerical complexity of the model discovery problem by exclusively including terms of known physical importance; see, e.g., \citeA{ReinboldKageorgeSchatzEtAl2021NatCommun} {\color{blue}for a discussion of fluid motivated examples}. For example, it may be suspected that the dynamics in a certain context are {\color{blue}driven by a background flow}, prompting the inclusion of {\color{blue}advective terms such as $\mathfrak{A}(\mathbf{u}_l)$} as per eq.~(\ref{eq:advection_operator}). In \S\ref{augmented} below, we discuss a strategy for representing library terms that are not integrable-by-parts{\color{blue}, like $\mathfrak{A}(\mathbf{u}_l)$,} in a weak form.

In the specific context of {\color{blue}atmospheric} fluid dynamics, it can be particularly helpful to consider a conservation perspective, noting that the {\color{blue}evolution} of a scalar {\color{blue}quantity} $\rho$ is given by the divergence of its flux $\boldsymbol{j}$, subject to any forcing $F$, \begin{align*}
    \rho_t = -\nabla\cdot\boldsymbol{j} + F(\boldsymbol{x},t,\rho, \, \dots).
\end{align*} For conserved quantities moving in tandem with a background velocity field $\boldsymbol{u}$, the flux is $\boldsymbol{j} = \rho\boldsymbol{u}$. An analogous {\color{blue}formulation for the flow} itself is that of the Euler equation, \begin{align*}
    \boldsymbol{u}_t = - \nabla\cdot\big(\boldsymbol{u} {\color{blue}\otimes} \boldsymbol{u} + p\mathbf{I}\big) + \boldsymbol{F}(\boldsymbol{x}, t, p, \boldsymbol{u}, \, \dots),
\end{align*} where $p$ represents an effective pressure. {\color{blue}In summary}, it often makes sense to include candidate terms $\{\mathcal{D}^if_j(\mathbf{U})\}$ that are cumulatively capable of representing any known or expected physical sources of flux and forcing.

In the present setting, one would like to select a physically-motivated library that is capable of parsimoniously representing transport phenomena such as {\color{blue}flux} $\nabla\cdot(\rho\boldsymbol{u})$, advection $(\boldsymbol{u}\cdot\nabla)\rho$, horizontal divergence $(\nabla\cdot\boldsymbol{u})\rho$, diffusion $\Delta \rho$, potential gradients $\nabla \Phi$, and forcing terms $F$ represented in terms of, e.g., quadratic monomials. {\color{blue}We note that \citeA{MessengerBortz2024IMAJNumerAnal} have shown that WSINDy converges to a similar class of models in the continuum data limit.} Since the forms that the relevant differential operators take depend on the coordinate system in which the original data $\mathbf{U}$ was measured, a library $\boldsymbol{\Theta}(\mathbf{U})$ is most useful when it is \lq{coordinate-aware.}' For example, in spherical surface coordinates $(\varphi, \theta)$ at $r=a$, additional metric terms appear in the differential operators, as in eqs.~(\ref{eq:advection_operator}) and (\ref{eq:divergence_operator}) above.  We also note that the Jacobian determinant in the discrete convolution of eq.~(\ref{eq:wsindy}) then becomes $a^2\cos(\theta) \Delta\varphi \Delta\theta$.

For the numerical experiments of \S\ref{simulated}, we adopt a standardized library {\color{blue}ansatz} to model individual scalar evolution equations of the form $\pt_t\mathbf{u}_l = \boldsymbol{\Theta}(\mathbf{U}) \mathbf{w}$, setting
\begin{align}\label{eq:standardizedLibrary}
    \boldsymbol{\Theta}(\mathbf{U}) =
    \begin{bmatrix}
    \, \vline & \vline &  & \vline & & \vline & \vline & \vline \\
    \, \mathbf{1} & {\color{blue}\mathcal{D}^1f_1(\mathbf{U})} & \cdots & \mathcal{D}^if_j(\mathbf{U}) & \cdots & {\color{blue}\mathcal{D}^If_J(\mathbf{U})} & \mathfrak{D}(\mathbf{u}_l) & \mathfrak{A}(\mathbf{u}_l) \\
    \, \vline & \vline &  & \vline & & \vline & \vline & \vline
    \end{bmatrix},
\end{align} where for each axis $x_i \in \{x_1, \, \dots, \, x_n\}$ and parameter $\alpha^i \in \{0, \, 1, \, 2\}$, we include a corresponding derivative $\mathcal{D}^i = \pt_{x_{i}}^{\alpha^i}$, along with the horizontal divergence and advection operators $\mathfrak{D}$ and $\mathfrak{A}$. For incompressible flows, we have $\mathfrak{D}(\mathbf{u})=\mathfrak{A}(\mathbf{u})$, in which case we do not duplicate the library column. {\color{blue}Beside a constant term $\mathbf{1}$}, we use a collection of $J$ functions $\{f_j\}_{j=1}^{J}$ that includes each possible quadratic monomial formed from pairwise components $\mathbf{u}_{k}${\color{red},} $\mathbf{u}_{l}$ of the discretized state variable $\mathbf{U}=[\mathbf{u}_1, \dots, \mathbf{u}_d]$, with $1 \leq k, \, l \leq d$, so that \begin{align*}
    f_j(\mathbf{u}_1, \, \dots, \, \mathbf{u}_d) \in \Big\{
    \mathbf{u}_1, \, \mathbf{u}_1^2, \, {\color{blue}\mathbf{u}_1\mathbf{u}_2}, \, \dots, \, {\color{blue}\mathbf{u}_k, \, \dots,} \, \mathbf{u}_k\mathbf{u}_{l}, \, \dots, \, {\color{blue}\mathbf{u}_d, \, \dots, \,} \mathbf{u}_{d}\mathbf{u}_{d-1},\, \mathbf{u}_d^2
    \Big\}.
\end{align*} {\color{blue}When two library columns $\Theta(i,j)$ and $\Theta(i',j')$ are nearly collinear over the data $\mathbf{U}$, we remove one column to ensure the condition number $\kappa(\mathbf{G})$ is tenable; see the appendix (\S\ref{dataset_descriptions}) for a detailed accounting of terms for each example.}

{\color{blue}We emphasize that one can also discover models posed in terms of spatial derivatives, as we do when discovering divergence models of the form $\pt_{x} \mathbf{u}_l = \boldsymbol{\Theta}(\mathbf{U})\mathbf{w}$. Due to the fact that spatial derivatives of a given atmospheric variable are commonly influenced by complicated and context-dependent stresses, it is difficult to prescribe a universal form of library for spatial \lq{evolution}' equations. Indeed, this very difficulty is a motivating factor for the study of symbolic identification of closure models from data in works like \cite{ZannaBolton2020GeophysicalResearchLetters}, and merits a substantial amount of future work. In any case, duplicate terms (e.g., $\pt_x \mathbf{u}_l$) should be removed from the ansatz library given in eq.~(\ref{eq:standardizedLibrary}) to prevent the discovery of trivial identities.}

\subsubsection{Augmented Libraries}\label{augmented}
Here, we present a strategy for creating an \textit{augmented library} $\boldsymbol{\Theta}'$ that implicitly includes the advection operator $\mathfrak{A}$, evaluated on the data, in a weak form. In instances of incompressible flow where $\nabla\cdot\boldsymbol{u}=0$, the divergence operator collapses to the advection operator, $\mathfrak{D}(\rho) = \nabla \cdot (\rho\boldsymbol{u}) = (\boldsymbol{u}\cdot\nabla)\rho = \mathfrak{A}(\rho)$. In these cases, the advection operator can be easily represented by including terms of the form $\mathcal{D}^i(\rho \boldsymbol{u}_j)$ in the library; e.g., in 2D Cartesian coordinates, we have $\mathfrak{A}(\rho) = \mathfrak{D}(\rho) = \pt_x(\rho u) + \pt_y(\rho v)$. However, in more general cases, some care is needed to represent differential operators {\color{blue}that are not integrable-by-parts} in a weak form -- that is, without resorting to the computation of pointwise derivatives from potentially noisy data.

To begin, we {\color{blue}assume} that we have access to horizontal divergence measurements $\{(\nabla \cdot \boldsymbol{u})(\boldsymbol{x}_m,t_m)\}_{m=1}^M$, as is the case with the assimilated data used in \S\ref{assimilated}; a discussion of {\color{red}a} more general approach is deferred to future work. Using the product rule for divergence, we consider the expansion $\mathfrak{D}(\rho) = \mathfrak{A}(\rho) + \mathfrak{C}(\rho)$, where $\mathfrak{C}(\rho) := (\nabla\cdot\boldsymbol{u})\rho$. As {\color{red}was} illustrated above for the case of 2D Cartesian coordinates, we assume that the {\color{blue}flux} $\mathfrak{D}(\mathbf{u}_l)$ admits a known expansion in terms of $N$ library columns $\Theta_{c_1}, \, \dots, \, \Theta_{c_{N}}$ given by \begin{align*}
    \mathfrak{D}(\mathbf{u}_l) = \sum_{n=1}^{N} {\color{blue}a_n} \Theta_{c_n}
    = \sum_{n=1}^{N} {\color{blue}a_n} \mathcal{D}^{i_n}f_{j_n}(\mathbf{U}),
\end{align*} where each column index $c_n \in \{1, \dots, IJ\}$ refers to a corresponding derivative index $i_n \in \{1,\dots,I\}$ and function index $j_n \in \{1,\dots,J\}$. The form of this expansion will depend on the coordinate system being used and the spatial dimension of the system. Similarly, we assume that the horizontal divergence data is given in the ${c_{N+1}}^{\rm{th}}$ column, with $\mathfrak{C}(\mathbf{u}) = a_{N+1}\Theta_{c_{N+1}}$. Given the {\color{blue}coefficients and} column indices {\color{blue}$\{(a_n, c_n)\}_{n=1}^{N+1}$}, we in turn define an augmented matrix $\mathbf{A} \in \mathbb{R}^{IJ\times(IJ+1)}$ via \begin{align*}
    \mathbf{A} := \begin{bmatrix}
        \ \mathbf{I}_{IJ\times{IJ}} & \vline & \!\!\mathbf{a} \ \
    \end{bmatrix},
    \quad \text{where} \quad
    \mathbf{a}_n := \begin{cases}
        a_n, \qquad \ \ \text{if $n \in \{c_1, \, \dots, \, c_{N}\}$},
        \vspace{-2mm}
        \\
        {\color{blue}-a_{N+1}, \ \ \text{if $n = c_{N+1}$},}
        \vspace{-2mm}
        \\
        0, \qquad \quad \ \, \text{otherwise}.
    \end{cases}
\end{align*} Right multiplication of $\mathbf{A}$ against the original library $\boldsymbol{\Theta}$ yields an augmented library $\boldsymbol{\Theta}'$ which includes the advection term; that is, $\boldsymbol{\Theta}' := \boldsymbol{\Theta} \mathbf{A} = [\boldsymbol{\Theta} \ \vline \ \mathfrak{D}(\mathbf{u}) - \mathfrak{C}(\mathbf{u})] = [\boldsymbol{\Theta} \ \vline \ \mathfrak{A}(\mathbf{u})]$. To arrive at an analogously augmented weak library, we simply define $\mathbf{G}' := \mathbf{G}\mathbf{A}$.

Importantly, we note that the technique of library augmentation is not limited to advection operators. In fact, any operator $\mathfrak{F}(\mathbf{u}_l) \in \text{span}\{\Theta_1, \dots, \Theta_{IJ}\}$ is representable in this manner, {\color{blue}subject to} the caveat that this process induces collinearity among the corresponding library columns and can result in {\color{blue}singular matrices $\mathbf{G}$ if the necessary linearly dependent columns are not subsequently deleted}. To {\color{blue}help address issues relating to poor condition numbers} $\kappa(\mathbf{G})$, we direct the reader to the appendix (\S\ref{scale_invariance}), where we review a helpful preconditioning approach formulated by \citeA{MessengerBortz2021JComputPhys}.

\subsection{Performance Metrics}\label{metrics}
To help gauge the quality of the results, we report the coefficient of determination $R^2$ corresponding to each WSINDy regression, which is defined by \begin{align}\label{eq:rsquare}
    R^2 = 1 - \frac{|\!| \, \mathbf{r} \, |\!|_2^2}{\big|\!\big| \, \mathbf{b} - \overline{\mathbf{b}} \, \big|\!\big|_2^2},
\end{align} where $\mathbf{r} := \mathbf{b} - \mathbf{G}\mathbf{w}^{\star}$ is the {\color{blue}pointwise} residual vector and $\overline{\mathbf{b}} := (K{\color{red}d})^{-1} (\sum_{k{\color{red}l}} b_{k\color{red}{,l}}) \mathbf{1}$. This metric, which equals the proportion of the variance of $\mathbf{b}$ that is explained by the discovered sparse model $\mathbf{G}\mathbf{w}^{\star}$, satisfies $R^2 \leq 1$, with the values closer to 1 indicating a better performing model. Following \citeA{MessengerBortz2021JComputPhys}, we also report the normalized $\ell^{\infty}$ coefficient error, $E_{\infty}$, whenever the true model and its coefficients $\mathbf{w}^{\rm{true}}$ are known (i.e., in \S\ref{spherical} and \S\ref{barotropic}), given by $E_{\infty} := \max_{j} |w^{\star}_j - w_j^{\rm{true}}| / |w_j^{\rm{true}}|$. The $E_{\infty}$ coefficient error represents the maximum element-wise relative error incurred by the discovered model. Note that only terms with nonzero coefficients ${\color{red}w(i,j)}$ are considered {\color{red}to be} \lq{discovered}.'

To assess the extent to which the identified models recover the correct terms, we follow \citeA{LagergrenNardiniLavigneEtAl2020ProcRSocA} and \citeA{MessengerDwyerDukic2024JRSocInterface} in reporting the \textit{true positive ratio} (or \textit{Jaccard index}), defined by 
\begin{align*}
\text{TPR} = \frac{\text{TP}}{\text{TP} + \text{FP} + \text{FN}}.
\end{align*} Here, $\text{TP} = |\text{supp}(\mathbf{w}^{\star}) \, \cap \, \text{supp}(\mathbf{w}^{\rm{true}})|$ denotes the number of terms that were correctly identified as nonzero, $\text{FP} = |\text{supp}(\mathbf{w}^{\star}) \cap \text{supp}(\mathbf{w}^{\rm{true}})^{\rm{C}}|$ denotes the number of terms that were falsely identified as nonzero, and $\text{FN} = |\text{supp}(\mathbf{w}^{\star})^{\rm{C}} \cap \text{supp}(\mathbf{w}^{\rm{true}})|$ denotes the number of terms that were falsely identified as zero. Note that a TPR of 1 means that the true model has been discovered in its entirety {\color{red}while} a TPR of 0 {\color{red}indicates} that none of the correct terms were identified.

\subsubsection{Forecasting with the Discovered Model}\label{forecasting_description}
{\color{blue}
For data obtained via numerical simulation (i.e., the \lq{Spherical},' \lq{Barotropic},' and \lq{Stratified}' examples listed in \S\ref{spherical} through \S\ref{stratified}), we truncate the original dataset $\mathbf{U}$ temporally, keeping only snapshots taken over times $t \in [0, \tau]$ for $\tau < T$, and treat the resulting trimmed dataset as the \lq{training dataset}' accessible to the WSINDy model. In turn, we produce an extrapolated \lq{forecast dataset}' $\mathbf{U}^+$ by numerically integrating the discovered model forward in time for $t \in [\tau, \, T]$. To gauge the predictive power of the identified PDE, we then compare $\mathbf{U}^{+}$ to the corresponding output of ground-truth model, $\mathbf{U}^{\star}$, and list the results in Table~\ref{table:forecastingTable} (also see Figures~\ref{fig:spherical_forecast} and \ref{fig:barotropic_forecast}).
}

To {\color{blue}assess} the results of the forecasts, we examine the spatially-averaged relative error at time $t$, defined by \begin{align*}
    {\color{blue}\overline{\mathcal{E}}(t) 
    := \frac{\big(\overline{\mathbf{U}^{\star} - \mathbf{U}^{+}}\big)(t)}{\overline{\mathbf{U}^{\star}}(t)},}
    \quad \text{where} \quad
    \overline{\mathbf{U}}(t) 
    := \frac{1}{\color{red}M} \sum_{m=1}^{M} {\color{blue}\big|} \mathbf{U}(\boldsymbol{x}_m, t) {\color{blue}\big|}.
\end{align*} The relative error at the final time of the forecast is then defined by $\overline{\mathcal{E}}_F := \overline{\mathcal{E}}({\color{red}T})$. Likewise, we define the \textit{time until tolerance}, $t_{\rm{tol}}$, as first time $t$ (if such a time exists) such that $\overline{\mathcal{E}}(t)$ exceeds a $10\%$ threshold; that is, $t_{\rm{tol}} := \min \left\{t : \overline{\mathcal{E}}(t) \geq 0.1\right\}$. {\color{blue} We report the {\color{blue}non-dimensionalized} ratio $t_{\rm{tol}}/T_0$, where $T_0$ is the integral timescale of the system, {\color{red} which approximates the timescale over which fluid fluctuations remain temporally correlated}; see \S\ref{sec:forecast_details} for details.} {\color{red}The ratio $t_{\rm{tol}} / T_0$ thus measures the proportion of the timescale over which the forecast remains 90\% accurate, on average (i.e., a larger ratio is better).} To give a sense of the relative $L^2$ error {\color{blue}incurred during the entire forecast}, we also report the normalized RMS error {\color{red}incurred over the entire space-time domain}, defined as \begin{align*}
    \text{RMSE}
    := \frac{\left|\!\left| \mathbf{U}^{\star} - \mathbf{U}^{+}\right|\!\right|_{\rm{RMS}}}{\left|\!\left| \mathbf{U}^{\star}\right|\!\right|_{\rm{RMS}}}
    =
    \frac{\left|\!\left| \mathbf{U}^{\star} - \mathbf{U}^{+}\right|\!\right|_2}{\left|\!\left| \mathbf{U}^{\star}\right|\!\right|_2}.
\end{align*} In Table~\ref{table:forecastingTable}, the $R^2$ value is defined as in eq.~(\ref{eq:rsquare}) above, with the exception of replacing $\mathbf{r} \mapsto \mathbf{U}^{\star} - \mathbf{U}^{+}$ and $\mathbf{b} \mapsto \mathbf{U}^{\star}$. {\color{red}Here, the $R^2$ metric may be viewed as assessing the ability of the forecast to capture small-scale variations in the data (i.e., larger $R^2$ values are better).} We discuss the forecast results in \S\ref{forecasting}.

\subsubsection{Robustness to Noise}\label{robustness}
To illustrate the performance of WSINDy in the presence of noisy data $\mathbf{U} + \boldsymbol{\epsilon}$, we run repeated numerical experiments on corrupted versions of the \lq{Spherical}' (\S\ref{spherical}) and \lq{Barotropic}' (\S\ref{barotropic}) datasets (the two examples where the true coefficients are known) and report the resulting $E_{\infty}$ and TPR averages (see Figure~\ref{fig:numerical_results_1}). Following \citeA{MessengerBortz2021JComputPhys}, we add distinct realizations of artificial i.i.d. noise $\epsilon \in \mathcal{N}(0,\sigma^2)$ in a pointwise fashion to each element of $\mathbf{U}$, which {\color{red}are} computed by enforcing a variance $\sigma^2$ satisfying $\sigma = \sigma_{\rm{NR}} \lvert\!\lvert \mathbf{U} \rvert\!\rvert_2$ so that $\sigma_{\rm{NR}} = \lvert\!\lvert \boldsymbol{\epsilon} \rvert\!\rvert_2 / \lvert\!\lvert \mathbf{U} \rvert\!\rvert_2$, where $|\!| \cdot |\!|_2$ represents the (vectorized) $\ell^2$ norm. For both examples, we explore noise ratios $\sigma_{\rm{NR}}$ in the range $0 \leq \sigma_{\rm{NR}} \leq 1$ (i.e., up to $100\%$ of the magnitude of the data), generating $\sim 400$ artificially corrupted datasets by running {\color{blue}8 to 11} experiments at each of 40 incremented noise levels given by $\sigma_{\rm{NR}} \in \{0.025 \cdot j\}_{j=1}^{40}$. We discuss the results of these experiments in \S\ref{simulated}.

\newpage

\section{Numerical Results}\label{results}
We conduct numerical experiments in two contexts of increasing difficulty, using WSINDy to learn equations for simulated (\S\ref{simulated}) and assimilated (\S\ref{assimilated}) {\color{red}atmospheric data}. The assimilated data are produced by fitting high-resolution continuous fields to a sparse set of empirical {\color{red}meteorological} observations. In practice, the sparsity of such {\color{red}data} makes assimilation a ubiquitous procedure. Results obtained with simulated data provide a sense of the method's performance in optimal conditions. Conversely, the numerical experiments performed using assimilated data are intended to more closely represent many real-world applications, where a ground-truth model may not be known in its entirety.

Beside providing a comparison between simulated and assimilated data settings, the examples in \S\ref{simulated} and \S\ref{assimilated} were specifically chosen to test the ability of WSINDy to accurately discover governing equations for data {\color{blue}exhibiting} important meteorological phenomena: unstable jets in a viscous fluid (\S\ref{spherical}), barotropic turbulence (\S\ref{barotropic}), temperature transport (\S\ref{stratified}), and potential vorticity conservation (\S\ref{era5_description}). {\color{red}Together, these mechanisms are important in the formation and dynamics of large-scale weather patterns such as jet streams and extratropical cyclones \cite{Mcwilliams1984JFluidMech, KoolothSmithStechmann2022GeophysicalResearchLetters}.} \citeA{MessengerBortz2021JComputPhys} originally demonstrated that WSINDy can reliably identify models from data exhibiting a number of canonical physical mechanisms, including \lq\lq{spatiotemporal chaos, nonlinear waves, nonlinear diffusion, shock [waves],}'' and \lq\lq{complex limit cycles}.'' Our experiments thus aim to complement those of Messenger and Bortz in the context of atmospheric fluid dynamics.

\subsection{Implementation Details}\label{implementation_details}
Algorithmically, we initialize our WSINDy hyperparameters in accordance with{\\}\citeA{MessengerBortz2021JComputPhys}, who detail methods for selecting test function spectra $|\hat{\psi}|$ based on the data such that $\psi$ becomes an implicit noise filter. {\color{blue}In particular}, we use separable test functions $\psi(\boldsymbol{x},t) = \phi_t(t)\Pi_{i=1}^n\phi_i(x_i)$ supported on a discrete grid defined by $\boldsymbol{\ell}$, where each $\phi_i$ {\color{blue}for $x \in [-\ell_i\Delta_i, \, \ell_i\Delta_i]$} is given by \begin{align*}
    \phi_i(x) = \left[1-\frac{x^2}{\ell_{i}^2\Delta_i^2} \right]^{p_i}
    \quad \text{with} \quad
    p_i 
    := \max\left\{ \left\lceil
    \frac{\ln(\tau_0)}{\ln\big((2\ell_i-1)/\ell_i^2\big)}
    \right\rceil, \ \bar{\alpha}^i + 1 \right\}.
\end{align*} Here, the parameter $\tau_0 = 10^{-10}$ is a tolerance indicating the maximum allowable value of the discretized test function $\psi({\color{red}\mathcal{X}_{\Delta} \times T_{\Delta}})$ on boundary of its support, while $\bar{\alpha}^i$ is the maximum order of derivative taken with respect to $x_i$ or $t$.

{\color{blue} Although we do not investigate other choices of test functions here, we note that, in principle, the specific family of test functions being used has some bearing upon the numerical results obtained. The modern wisdom states that the two most important properties of a generating test function $\psi$ are: (1) its spectrum $|\hat{\psi}|$ (i.e., its bandwidth), which is determined by $\boldsymbol{\ell}$ and is responsible for setting its properties as a filter, and (2) its smoothness class $C_c^p$, which determines the numerical truncation error of the weak form integrals through the Euler-Maclaurin formula. In our case, the chosen separable components $\phi_i$, which are equivalent to rescaled Bernstein polynomials of degree $p_i$, induce a $\mathcal{O}(\Delta_i^{p_i+1})$ truncation error when used in tandem with the trapezoidal rule. Interestingly, these specific $\phi_i$ have enjoyed a long history, being used for similar purposes within the Modulating Function Method as early as the mid-1960's  \cite{LoebCahen1965IEEETransAutomControl}.}

We use the modified sequential thresholding least squares (MSTLS) routine {\color{blue}described in \cite{MessengerBortz2021JComputPhys}} to compute the model weights $\mathbf{w}^{\star}$ by minimizing a {\color{blue}normalized} loss function of the form of eq.~(\ref{eq:sindyLoss}), \begin{align}\label{eq:wsindy_loss}
    \mathcal{L}({\color{red}\mathbf{w}}) = \mathcal{L}\left(\mathbf{w}; \, \frac{\mathbf{b}_{\textsc{ls}}}{|\!| \mathbf{b}_{\textsc{ls}} |\!|_2}, \, \frac{\mathbf{G}}{|\!| \mathbf{b}_{\textsc{ls}} |\!|_2}\right)\!,
\end{align} where the Lagrange multiplier in eq.~(\ref{eq:sindyLoss}) is set to ${\color{red}\mu}=(IJ)^{-1}$ throughout. Here, $\mathbf{b}_{\textsc{ls}}$ is the ordinary least-squares estimate defined by \begin{align*}
    \mathbf{b}_{\rm{LS}} := \mathbf{G}\mathbf{w}_{\textsc{ls}},
    \quad \text{where} \quad
    \mathbf{w}_{\textsc{ls}} = {\color{blue}\left(\mathbf{G}^T\mathbf{G}\right)^{-1}\mathbf{G}^T}\mathbf{b}.
\end{align*} Note that \citeA{MessengerBortz2021JComputPhys} use the notation $\hat{\lambda}$ to denote the optimal threshold value used in the MSTLS algorithm, which is an {\color{blue}analogous but distinct} quantity from the Lagrange multiplier ${\color{red}\mu}$ as used here. {\color{blue}In particular, the MSTLS weights $\mathbf{w}^{\star}$ are obtained in accordance with a dominant balance rule of the form $\hat{\lambda} \! \leq \! |\!|w_i\mathbf{G}_i|\!|_2 / |\!|\mathbf{b}|\!|_2 \! \leq \! \hat{\lambda}^{-1}$; on a given iteration of MSTLS, weights $(\mathbf{w}_{\textsc{ls}})_i$ outside of this range are thresholded to $0$ and the least-squares problem is resolved over the remaining non-zero weights. To find the \lq{optimal}' threshold $\hat{\lambda} \in {\color{red}[0,1]}$, the algorithm scans over a set of candidate values $\{\lambda_i\}$ with $\log_{10}(\lambda_i)$ equally spaced from $-4$ to $0$, settling on a threshold value which corresponds to the weights $\mathbf{w}^{\star}$ minimizing $\mathcal{L}\color{red}(\mathbf{w})$ as specified in eq.~(\ref{eq:wsindy_loss}).}

In each example, we use a uniformly-spaced grid of query points $\{(\boldsymbol{x}_k, t_k)\}$, {\color{blue} the number of which we list in the appendix (see Table~\ref{table:dataset_descriptions})}. To alleviate numerical errors incurred due to high condition numbers $\kappa(\mathbf{G})$, we extend the scale-invariant preconditioning method of \citeA{MessengerBortz2021JComputPhys} to work with data of the form of eq.~(\ref{eq:standardizedLibrary}); see the appendix (\S\ref{scale_invariance}) for a description of this approach.

\begin{table}
\begin{center}
\begin{tabular}{||c c c c c c||}
\hline
\textbf{Dataset} & \textbf{Discovered PDE} & $\boldsymbol{R^2}$ \textbf{(\%)} & $\boldsymbol{\mathcal{L}(\mathbf{w}^{\star})}$ & $\boldsymbol{E_{\infty}}$ & \textbf{TPR} \\ [1ex]
\hline\hline
Spherical & $h_t = -\nabla\cdot(h\boldsymbol{u}) - H_0(\nabla\cdot\boldsymbol{u}),$ & 100 & 0.04 & 8.1e-4$^\dagger$ & 1$^\dagger$ \\ [1ex]
& $u_t = -(\boldsymbol{u}\cdot\nabla)u - fv - g_1 (\nabla{h})_1,$ & 100 & 0.09 & 8.3e-4$^\dagger$ & 1$^\dagger$ \\ [1ex]
& $v_t = -(\boldsymbol{u}\cdot\nabla)v + fu - g_2 (\nabla{h})_2$ & 100 & 0.08 & 2.3e-4$^\dagger$ & 1$^\dagger$ \\ [1ex]
& $H_0 = 1.57\text{e-}3, \ g_1 = 19.96, \ g_2 = 19.94$ & & & & \\ [1ex]
\arrayrulecolor{gray}\hline\arrayrulecolor{black}
Barotropic & $\zeta_t = -\nabla\!\cdot(\zeta\boldsymbol{u})$, & 100 & {\color{red}0.03} & 2.6e-3$^\dagger$ & 1$^\dagger$ \\ [1ex]
& ${\color{red}u_x = -v_y}$ & 100 & {\color{red}0.03} & 3.0e-12 & 1 \\ [1ex]
\arrayrulecolor{gray}\hline\arrayrulecolor{black}
Stratified & $\vartheta_t = - \alpha_1 \left(\vartheta^2\right)_x - \alpha_2 \left(\vartheta^2\right)_y - \beta \vartheta_y$ & 94.9 & 0.18 & \textsc{n/a} & \textsc{n/a} \\ [1ex]
& $\alpha_1 = 2.80, \ \alpha_2 = 2.53, \ \beta = 3.11$ & & & & \\ [1ex]
\hline
\end{tabular}
\end{center}
\caption{Model identification results for the simulated datasets of \S\ref{simulated} at $0\%$ noise. {\color{red}The metrics are defined in \S\ref{metrics} while the datasets are detailed in the appendix (\S\ref{dataset_descriptions}).} For {\color{red}legibility}, each value is rounded to two decimal places, except for the $R^2$ values, which are rounded to three. The $^\dagger$ denotes that a numerical stability term was neglected (see \S\ref{spherical}, \S\ref{barotropic} for details).}
\label{table:simulated}
\end{table}

\subsection{Simulated Data}\label{simulated}
Numerical results from the \lq{Spherical}' dataset (\S\ref{spherical}) illustrate the application of WSINDy to large-scale geophysical flows in a spherical coordinate system. Given \textit{direct} measurements of the relevant physical forces ($\mathfrak{A}(h), \boldsymbol{f}$, etc.), we find that WSINDy correctly identifies the shallow water equations governing a buoyant fluid surface $h$ and the horizontal velocity $\boldsymbol{u}$ on the surface of a sphere (see Table~\ref{table:simulated}), with performance metrics largely mirroring those of the \lq{Barotropic}' dataset. We note that the hyperviscosity terms ({\color{red}i.e.,} $\nu\Delta^2(\, \cdot \,)$ in \S\ref{spherical}, where $\nu \sim 10^{-9}$) are not considered {\color{red}to be} true terms in the model, since they are added for numerical stability purposes.  Remarkably, we find that for noise-less data ($\sigma_{\rm{NR}} = 0$), WSINDy identifies models with coefficients of determination on the order ${\color{red}R^2 \sim 0.995}$ for both the \lq{Barotropic}' and \lq{Spherical}' datasets, indicating that the discovered models {\color{red}account for} roughly \textit{all} of the variance of the data. {\color{red}Moreover, no spurious terms were identified for any of the examples listed in Table~\ref{table:simulated} (i.e., ${\rm{FP}} = 0$).}


{\color{blue}Moving} to the \lq{Barotropic}' (\S\ref{barotropic}) section of Table~\ref{table:simulated}, we observe that the correct models for both the vorticity $\zeta$ and divergence $\nabla \cdot \boldsymbol{u}$ are identified from simulations of highly turbulent barotropic air flow. Moreover, the upper panels of Figure~\ref{fig:numerical_results_1} demonstrate that this performance is robust to the presence of high-magnitude i.i.d. noise in the data. Notably, incrementally increasing the noise ratio $\sigma_{\rm{NR}}$ from $0$ (i.e., noiseless data) to $1$ (i.e., noise of equal magnitude to the data) {\color{blue}tends to affect only} the $E_{\infty}$ coefficient error, with the identified PDE often maintaining the correct selection of terms. In particular, WSINDy recovers the correct form of the vorticity equation ($\rm{TPR} = 1$) in each trial where $\sigma_{\rm{NR}} \leq 0.825$. For noise levels $\sigma_{\rm{NR}} > 0.825$, WSINDy misidentifies spurious terms in roughly $5\%$ to $10\%$ of the corresponding trials; again, see Figure~\ref{fig:numerical_results_1}.

The \lq{Stratified}' example (\S\ref{stratified}) illustrates the recovery of an effective PDE model that describes the dominant physical mechanisms in the data -- in this case, a {\color{red}nonlinear} traveling wave model of the form $\vartheta_t = - {\color{red}\alpha_1} \left(\vartheta^2\right)_x - {\color{red}\alpha_2} \left(\vartheta^2\right)_y - {\color{red}\beta} \vartheta_y$. {\color{red}To help interpret this entry of Table~\ref{table:simulated}, we refer the reader to} \citeA{RudyBruntonProctorEtAl2017SciAdv}, {\color{red}who} provide an interesting discussion of the discovery of linear versus nonlinear wave equations in the context of solitary and interacting {\color{red}Korteweg-de Vries} solitons. Although in this instance the true form of the coefficients $\mathbf{w}^{\rm{true}}$ are \textit{not} exactly known (see \S\ref{stratified}), we note that a purely-advective PDE is an especially plausible result, given that the model is driven by a constant geostrophic wind $\boldsymbol{v}_g$ (see Figure~\ref{fig:geophysical_wave}). Moreover, this effective transport equation is found to describe roughly $94.9\%$ of the variance of the data.

Interestingly, in the \lq{Barotropic}' {\color{red}example}, the average $E_{\infty}$ error is \textit{not} observed to initially increase with the noise level $\sigma_{\rm{NR}}$ and instead obtains a local minimum near {\\}$\sigma_{\rm{NR}} \approx 0.15$ (see Figure~\ref{fig:numerical_results_1}, upper-left panel). This phenomenon appears to be related to \textit{stochastic resonance}, in which the power of a signal is effectively boosted with the addition of  noise. We {\color{blue}offer} a conjecture {\color{blue}as to} the cause of this apparent resonance. In particular, we note that a sub-grid dissipation model, $\texttt{ssd}$, is implemented in eq.~(\ref{eq:pyqg}), which is achieved by filtering out high-frequency content ${\color{red}\mathbf{u}' \mapsto \mathbf{u}}$ in the spectrum $|\hat{\mathbf{u}}'|$ of the unfiltered data, {\color{blue}changing the effective dynamics}. Intuitively, one would expect the addition of Gaussian noise $\mathbf{u} \mapsto \mathbf{u} + \boldsymbol{\epsilon}$ to reintroduce high-frequency components back into the spectrum $|\hat{\mathbf{u}} + \hat{\boldsymbol{\epsilon}}|$. At some critical noise level $\sigma_{\rm{NR}} = \sigma_{\rm{crit}} \, {\color{red}> 0}$, we may obtain the closest match between the spectra, with $|\hat{\boldsymbol{u}} + \hat{\boldsymbol{\epsilon}}| \approx |\hat{\mathbf{u}}'|$. Since the unfiltered data $\mathbf{u}'$ {\color{red}follow} the non-dissipative model identified by WSINDy ({\color{blue}that is,} with $\texttt{ssd} = 0$), the {\color{red}listed} coefficient $w_i = 1$ of the advection term $(\boldsymbol{u}\cdot\nabla)\zeta$ would, in this case, be more {\color{red}a more} accurate {\color{red}representation} {\color{blue}near} $\sigma_{\rm{NR}} \approx \sigma_{\rm{crit}}$ than at $\sigma_{\rm{NR}} = 0$.

\begin{table}
\begin{center}
\begin{tabular}{||c c c c||} 
\hline
\textbf{Variable} & \textbf{Discovered PDE} & $\boldsymbol{R^2}$ \textbf{(\%)}& $\boldsymbol{\mathcal{L}(\mathbf{w}^{\star})}$ \\ [0.5ex] 
\hline\hline
$\omega$ & $\omega_t = -\alpha(\boldsymbol{u} \cdot \nabla)\omega + \beta\dot{\eta}\omega$ & 74.4 & 0.18 \\ [1ex]
& $\alpha = 0.70, \ \beta = 0.29\Omega$ & & \\ [1ex]
\arrayrulecolor{gray}\hline\arrayrulecolor{black}
$u$ & $u_t = -\alpha(\boldsymbol{u} \cdot \nabla)u - \beta(\nabla\cdot\boldsymbol{u})u + \mathcal{R}_1 + \mathcal{R}_2$ & 23.2 & 0.64 \\ [1ex]
& $\mathcal{R}_1 := \gamma\frac{\tan(\theta)uv}{a}$ & & \\ [1ex]
& $\mathcal{R}_2 := - [\delta\pt_{\varphi} + \mu\pt_{\theta}]\left(\frac{u}{a\cos(\theta)}\right) - \nu\pt_{\varphi}\left(\frac{u^2}{a\cos(\theta)}\right)$ & & \\ [1ex]
& $(\alpha, \beta, \gamma, \delta, \mu, \nu) = (0.06, \, 0.04. \, 0.41, \, 1.56, \, 0.26, \, 0.07)$ & & \\ [1ex]
\arrayrulecolor{gray}\hline\arrayrulecolor{black}
$v$ & $v_t = -\alpha(\boldsymbol{u} \cdot \nabla)v + \beta \frac{u}{a\cos(\theta)}$ & 30.7 & 0.36 \\ [1ex]
& $\alpha = 0.21, \ \beta = 2.37$ & & \\ [1ex]
\hline
\end{tabular}
\end{center}
\caption{ Model discovery results for the assimilated ERA5 data of \S\ref{assimilated} (cf. \S\ref{IFS_eqns}), which are detailed in \S\ref{era5_description}. Results are reported using the same rounding scheme as Table~\ref{table:simulated}.}
\label{table:assimilated}
\end{table}

\subsection{Assimilated Data}\label{assimilated}
We use assimilated {\color{red}meteorological} data from the from \citeA{CopernicusClimateChangeService2023} and the ECMWF Reanalysis v5 (ERA5), which implements the 4D variational (4D-Var) data assimilation algorithm described by \citeA{AnderssonThepaut2008ECMWFNewsl} (see \S\ref{era5_description}). In 4D-Var, an atmospheric state is estimated from sparse weather data by statistically interpolating between weighted empirical observations and an IFS forecast from an analysis performed 12 hours prior, minimizing a corresponding loss function. The forecast is generated with models similar to eqs.~(\ref{eq:IFSmomentum_u}) and (\ref{eq:IFSmomentum_v}) below.

A noteworthy result is that, given {\color{red}synoptic scale} assimilated data, WSINDy recovers a conservation law-like model governing the evolution of potential vorticity $\omega$ in the upper troposphere (see Table~\ref{table:assimilated}). The discovery that potential vorticity is conserved in adiabatic air flow is itself regarded as one of the important meteorological results of the $20^{\rm{th}}$ century; see \citeA{KoolothSmithStechmann2022GeophysicalResearchLetters} for a discussion. We note that the identified PDE, which takes the form $\omega_t + \alpha(\boldsymbol{u} \cdot \nabla)\omega - \beta\dot{\eta}\omega = 0$, differs from a true conservation law $\omega_t + \alpha(\boldsymbol{u} \cdot \nabla)\omega - \beta\dot{\eta}\omega_{\eta} = 0$ via the {\color{red}exclusion} of a vertical gradient term $\dot{\eta}\omega_{\eta}$. This is explained by noting that a single pressure level ($p = 200$ hPa) was used in the training data, precluding the construction of the vertical derivative $\omega_{\eta}$ and prompting the inclusion of a correlated term, $\omega$. This phenomenon also illustrates the influence of the query point placement (i.e., the choice of points $\{(\boldsymbol{x}_k, t_k)\}$), on the model discovery process.

To complement the potential vorticity model, Table~\ref{table:assimilated} also lists corresponding momentum equations, which were subsequently discovered at the same scales $\boldsymbol{\ell}$. While the potential vorticity model explains roughly $74.4\%$ of the variance of the data, the $u_t$ and $v_t$ models fare substantially worse at $23.2\%$ and $30.7\%$, respectively. Since the latter models take intuitively plausible forms, cf. eqs.~(\ref{eq:IFSmomentum_u}) and (\ref{eq:IFSmomentum_v}), this lack of agreement is presumably due to {\color{blue}either} the presence of large latitudinal variations in the data (e.g., the jet streams observed in Figure~\ref{fig:era5data}) {\color{blue}or to ill-conditioned $\mathbf{G}$ matrices}. We suspect that a more nuanced treatment of non-autonomous terms (varying with $\varphi$, $\theta$), vertical terms (varying with $\eta$), and thermodynamic terms (e.g., varying with $\rho$, $p$) in the candidate library would increase the descriptive capacity of the discovered models. {\color{blue}Moreover, we expect that an improved numerical implementation {\color{red}based upon a scale analysis} would dramatically increase the accuracy of the coefficient estimates.} {\color{red}Alternatively, future work could consider an approach like that of \citeA{ZannaBolton2020GeophysicalResearchLetters} and instead attempt to discover a parameterized \textit{closure} model of the form $\mathcal{D}^0\boldsymbol{u} = \mathcal{R}[\boldsymbol{u}]$, where $\mathcal{D}^0$ is an idealized PDE model such as IFS equations~(\ref{eq:IFSmomentum_u}) and (\ref{eq:IFSmomentum_v}) given below, and $\mathcal{R}$ is a `residual operator' analogous to the complicated spatial stresses $\mathcal{R}_1$ and $\mathcal{R}_2$ found in Table~\ref{table:assimilated}.}

\subsection{Distribution of Residuals}\label{residuals}
\begin{figure}[htb!]
    \centering
    \includegraphics[width=0.495\linewidth]{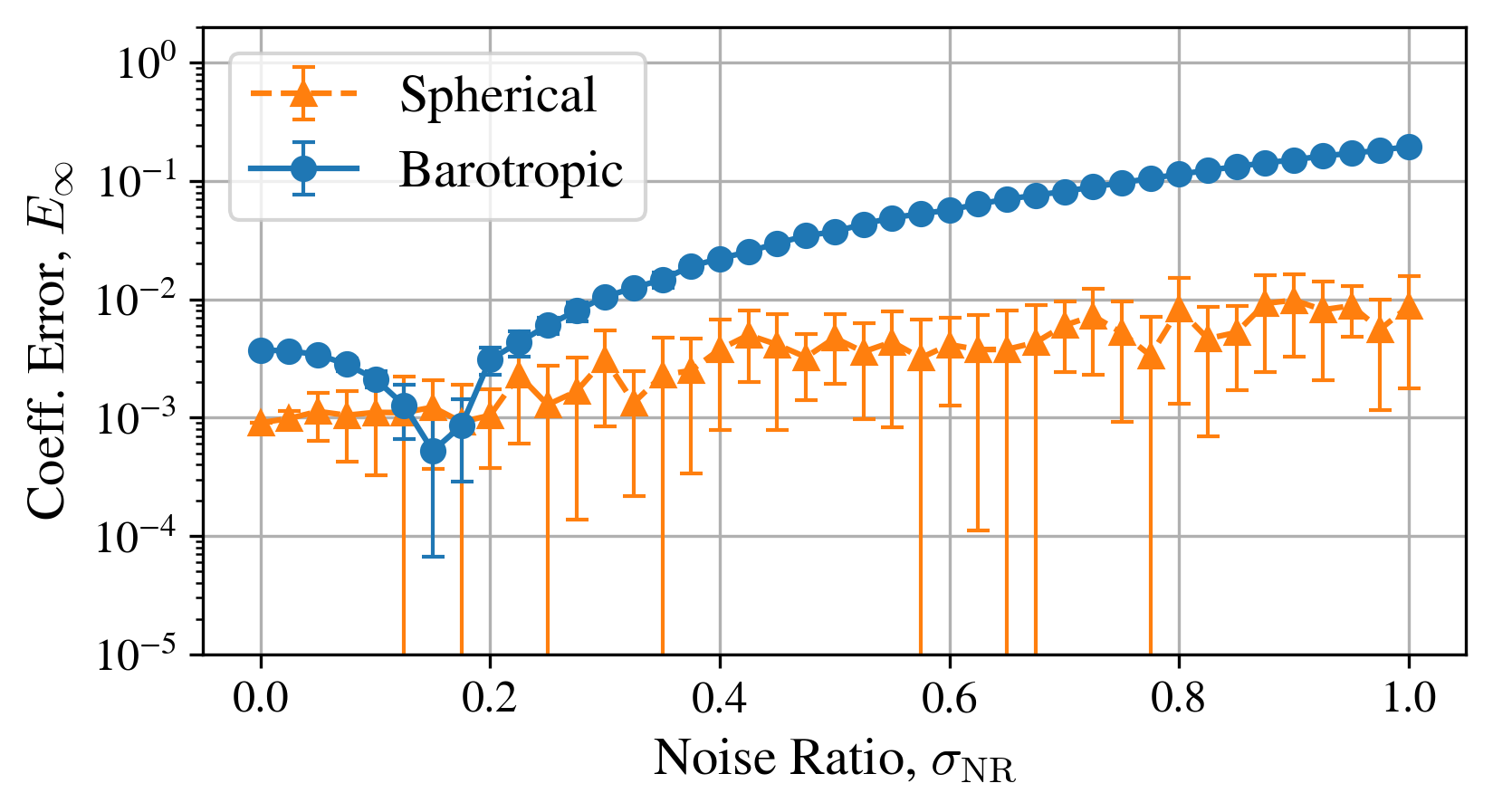}
    \includegraphics[width=0.495\linewidth]{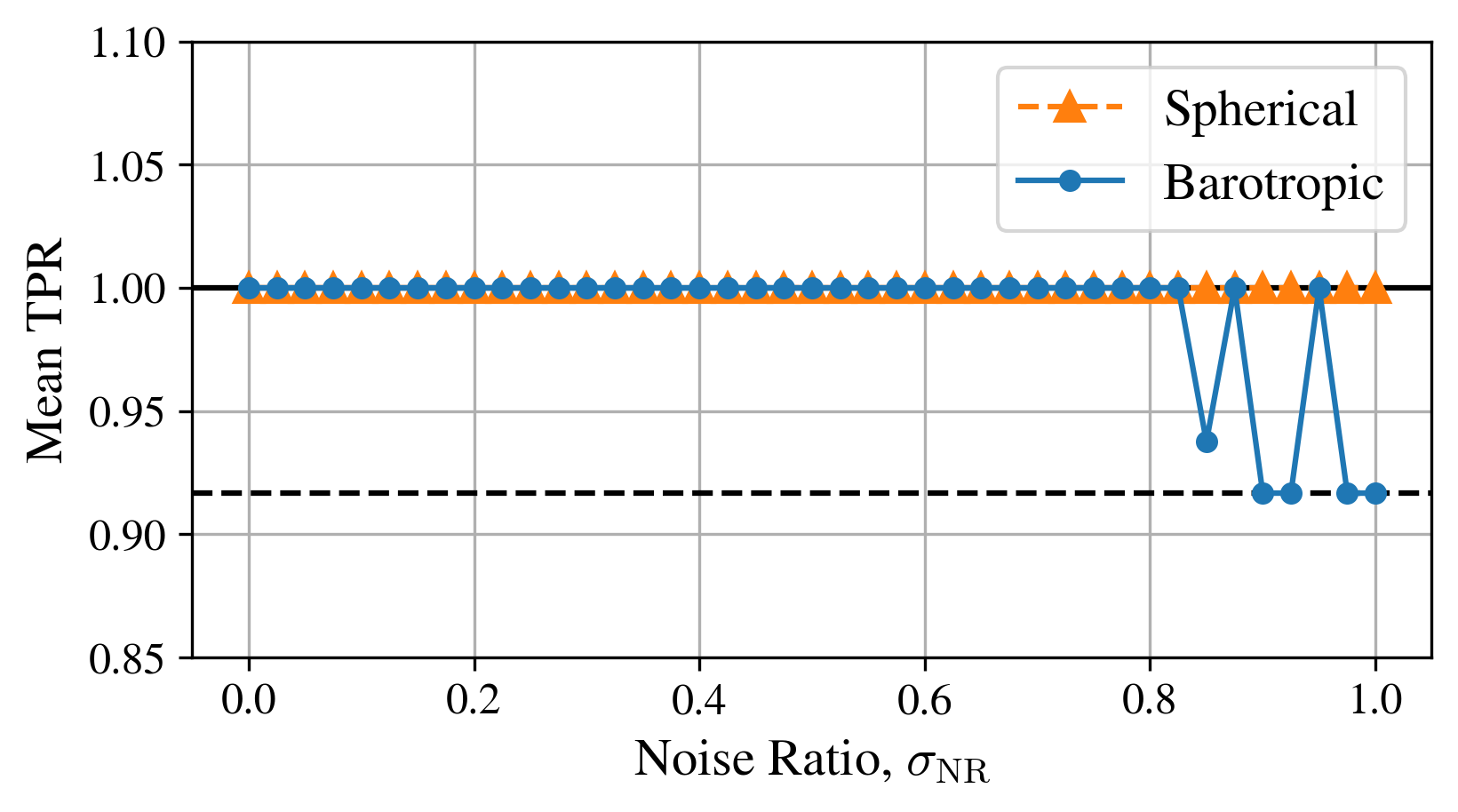}
    \includegraphics[width=0.495\linewidth]{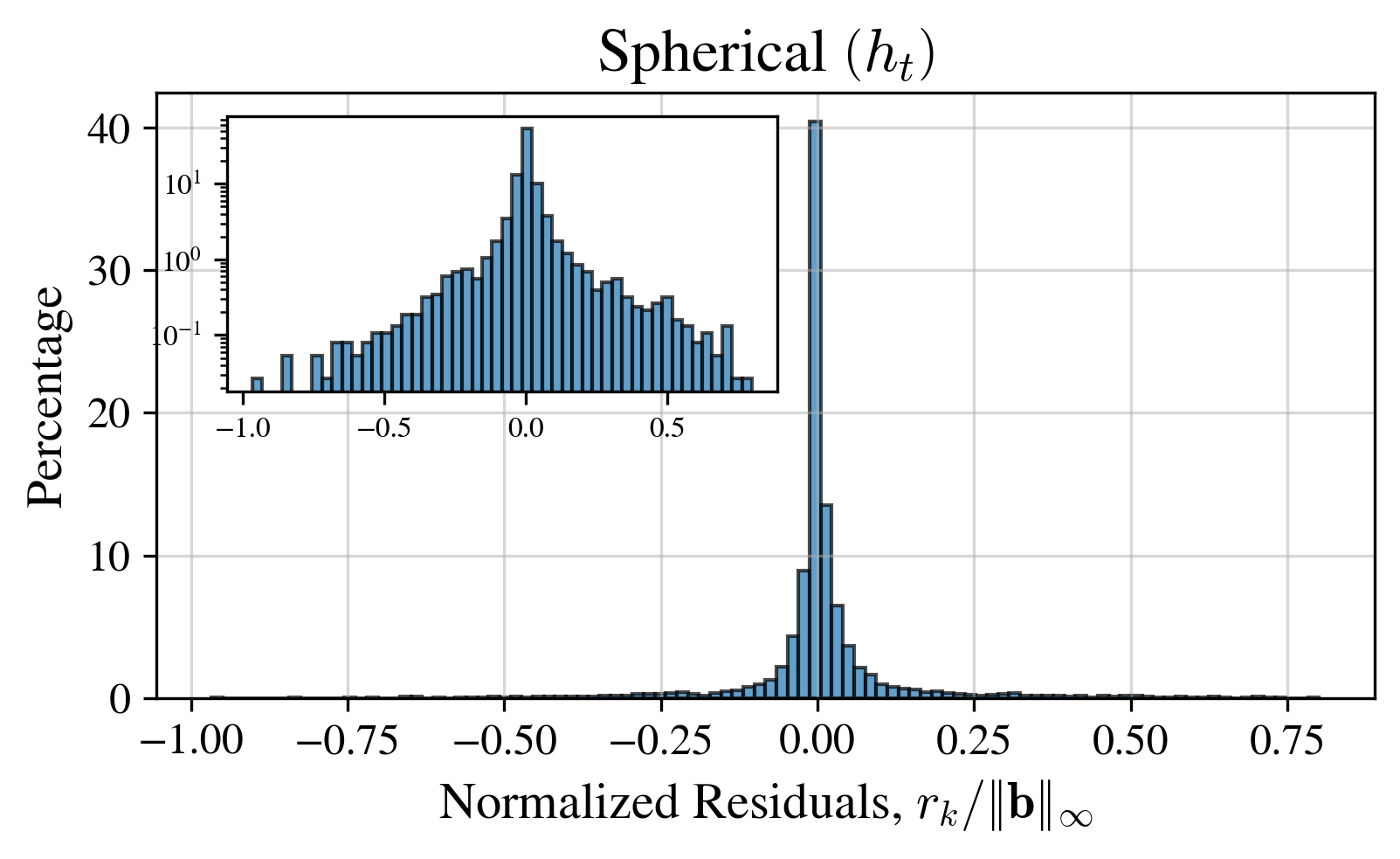}
    \includegraphics[width=0.495\linewidth]{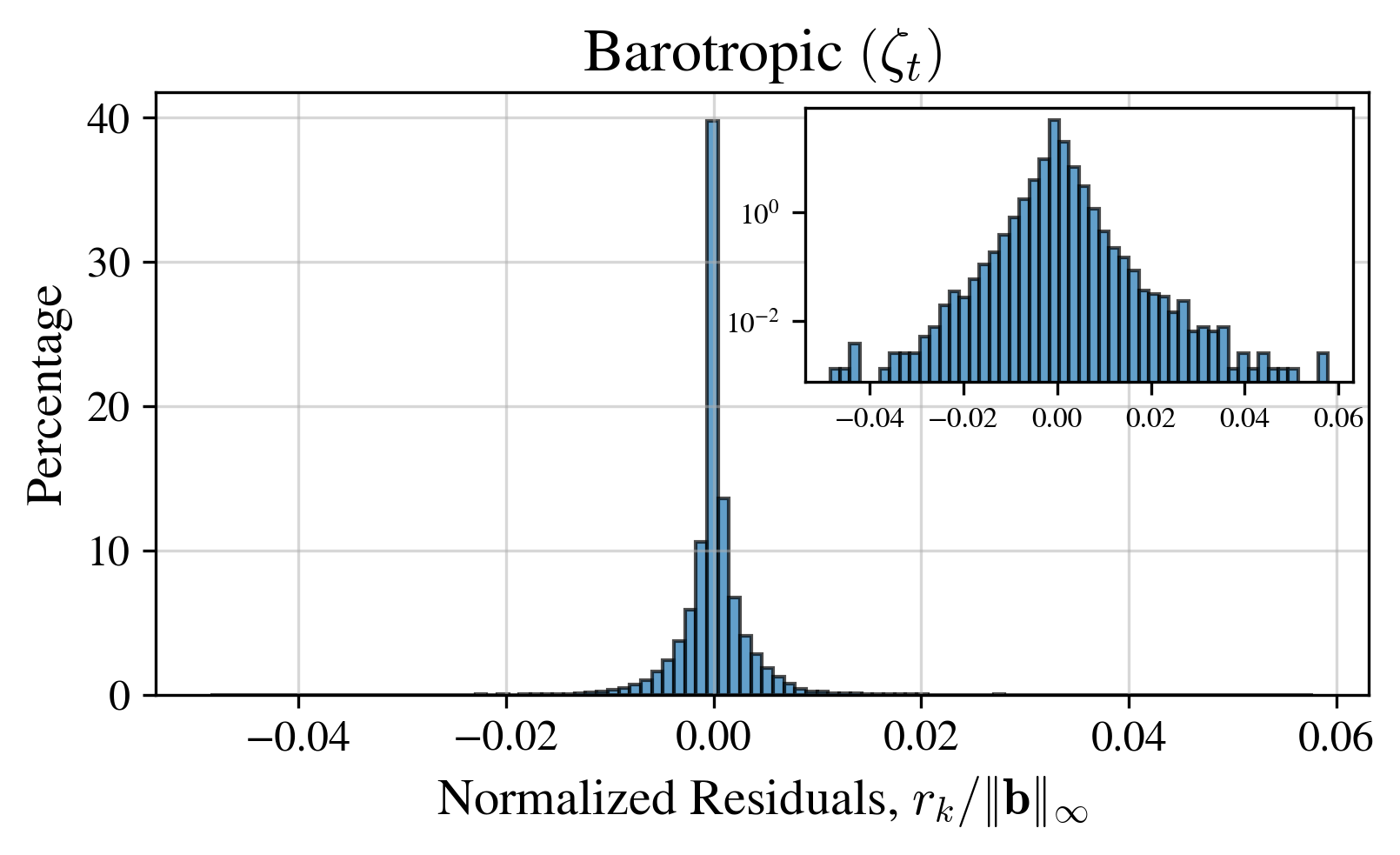}
    \includegraphics[width=0.495\linewidth]{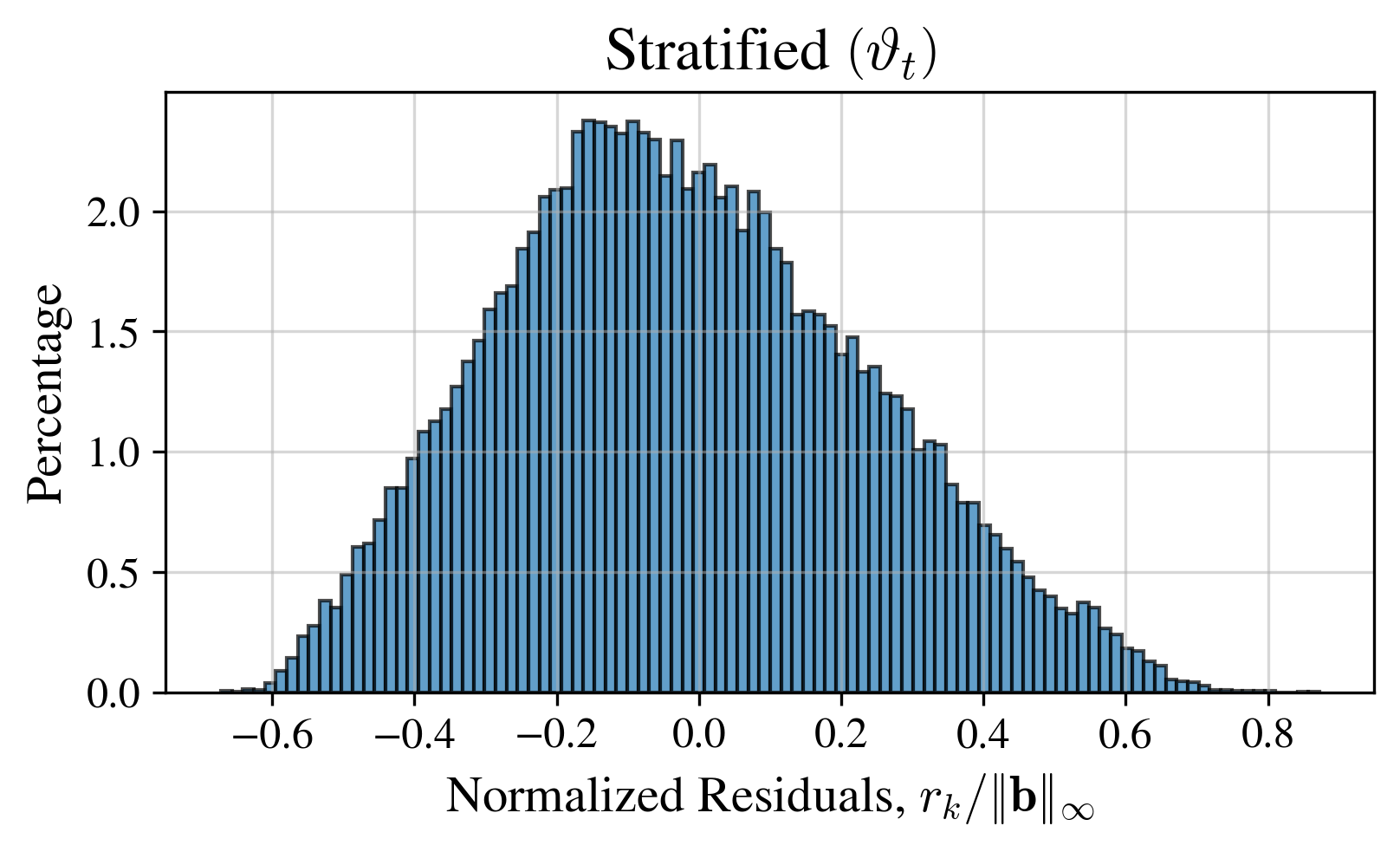}
    \raisebox{1.3mm}{\includegraphics[width=0.495\linewidth]{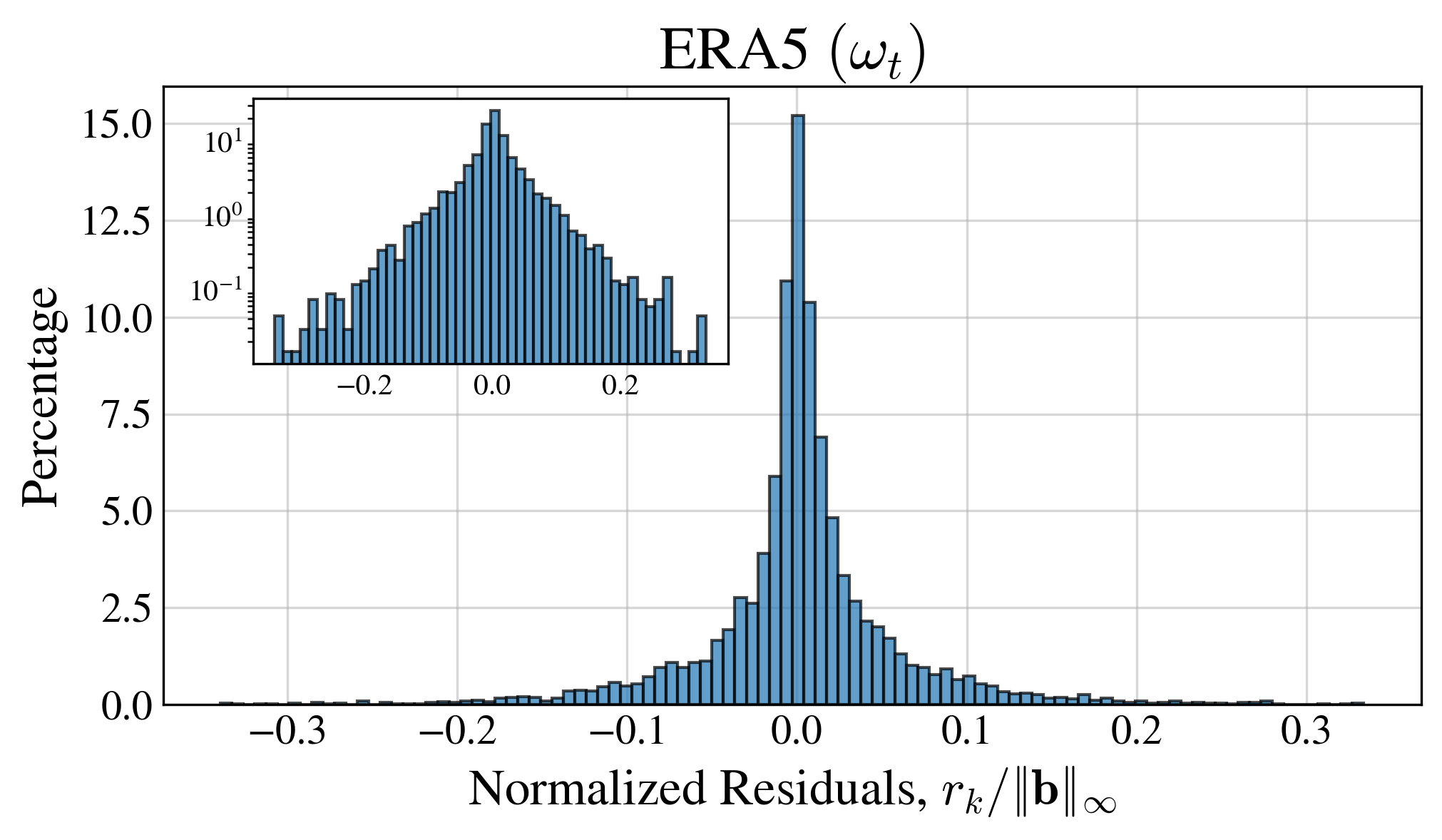}}
    \caption{Overview of numerical experiments with the \lq{Spherical}' and \lq{Barotropic}' datasets, where the true coefficients are known, {\color{blue}and the \lq{Stratified}' (\S\ref{stratified}) and \lq{ERA5}' (\S\ref{assimilated}) datasets, where the true coefficients are not known}. The top panels illustrate the mean $E_{\infty}$ error (top-left) within $1\sigma$ error bars and mean TPR (top-right), obtained using {\color{blue}8 to 11} realizations of Gaussian i.i.d. noise $\boldsymbol{\epsilon}$ at each noise level $\sigma_{\rm{NR}} \in \{0.025 \cdot j\}_{j=1}^{40}$. The bottom panels illustrate the distribution of residual values $\{r_k\}$ of eq.~(\ref{eq:wsindy}), where the residual vector is given by $\mathbf{r} = \mathbf{b} - \mathbf{G}\mathbf{w}^{\star}$. Each panel corresponds to a model identification result reported in Table~\ref{table:simulated} or Table~\ref{table:assimilated} at $0\%$ noise; (center-left) height $h$ from the \lq{Spherical}' dataset of \S\ref{spherical}, (center-right) vorticity $\zeta$ from the \lq{Barotropic}' dataset of \S\ref{barotropic}, {\color{blue}(bottom-left) potential temperature $\vartheta$ from the \lq{Stratified}' dataset, (bottom-right) potential vorticity $\omega$ from the \lq{ERA5}' dataset}. See \S\ref{simulated} and \S\ref{residuals} for a discussion, as well as Figure~\ref{fig:additional_residuals} in the appendix for additional residual plots.}
    \label{fig:numerical_results_1}
\end{figure}

In Figure~\ref{fig:numerical_results_1}, we plot component-wise histograms of the WSINDy {\color{red}equation} residuals $\mathbf{r} = \mathbf{b} - \mathbf{G}\mathbf{w}^{\star}$, where $\mathbf{G}$ and $\mathbf{b}$ are computed via eq.~(\ref{eq:wsindy}) at $0\%$ noise and $\mathbf{w}^{\star}$ is computed by applying MSTLS to the loss function $\mathcal{L}$ defined in eq.~(\ref{eq:wsindy_loss}). Of the four distributions displayed, only that of the \lq{Stratified}' dataset appears to be {\color{blue}even approximately} normally distributed, indicating the presence of correlated errors $\{r_k\}$. {\color{blue}In the case of} the \lq{Spherical}' dataset, {\color{blue}we expect that} that the non-normality {\color{blue}is due in part to} the underlying spatial anisotropicity of the data (see Figure~\ref{fig:spherical_forecast}). However, we note that the distributions corresponding {\color{blue}\textit{each} of the \lq{Spherical},' \lq{Barotropic}' and \lq{ERA5}' datasets} resemble product distributions (i.e., Bessel-K distributions) arising from, e.g., the application of ordinary least-squares to an errors-in-variables problem. These examples indicate that an iteratively-reweighted least-squares routine, such as the WENDy algorithm of \citeA{BortzMessengerDukic2023BullMathBiol}, may improve the corresponding parameter estimates. We note that the application of WENDy to the setting of PDEs is currently an area of active research.

\newpage

\newpage

\begin{figure}
    \centering
    \includegraphics[width=0.96\linewidth]{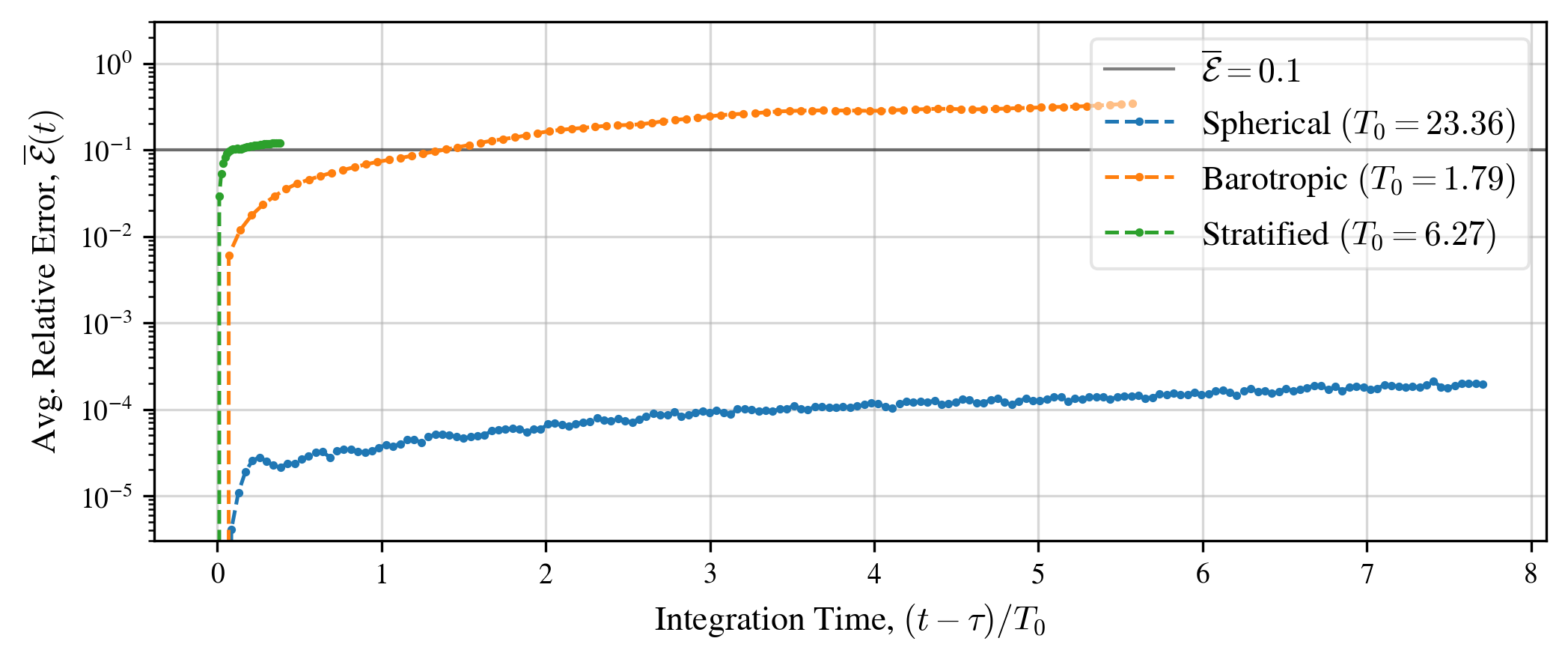}
    \caption{Forecasting error as a function of the non-dimensionalized integration time.}
    \label{fig:integral_timescales}
\end{figure}

\begin{table}
\begin{center}
\begin{tabular}{||c c c c c c||} 
\hline
\textbf{Dataset} & \textbf{Test Data, {\color{red}$\boldsymbol{[\tau,T]}$}} & \textbf{RMSE (\%)} & $\boldsymbol{R^2}$ \textbf{(\%)} & {\color{blue}$\boldsymbol{t_{\rm{tol}}/T_0}$} & $\boldsymbol{\overline{\mathcal{E}}_F}$ \textbf{(\%)} \\ [0.5ex] 
\hline\hline
Spherical ($h$) & ${\color{red}t \in [360, 540]}$ & 0.01 & 100 & \textsc{n/a} & 0.02 \\ [1ex]
Barotropic ($\zeta$) & ${\color{red}t \in [20, 30]}$ & 20.7 & 95.7 & {\color{red}1.39} & 34.5 \\ [1ex]
{\color{blue}Stratified ($\vartheta$)} & ${\color{red}t \in [5.1, 7.5]}$ & {\color{blue}13.6} & {\color{blue}46.5} & {\color{red}0.02} & {\color{blue}12.1} \\ [1ex]
\hline
\end{tabular}
\end{center}
\caption{Forecasting results reported using the metrics defined in \S\ref{forecasting_description}. Results are rounded to three decimal places, except for ${\color{blue}t_{\rm{tol}}/T_0}$, which is rounded to two. See \S\ref{forecasting} for a discussion.}
\label{table:forecastingTable}
\end{table}


\subsection{Forecasting Accuracy}\label{forecasting}
\begin{figure}
    \centering
    \includegraphics[width=\linewidth]{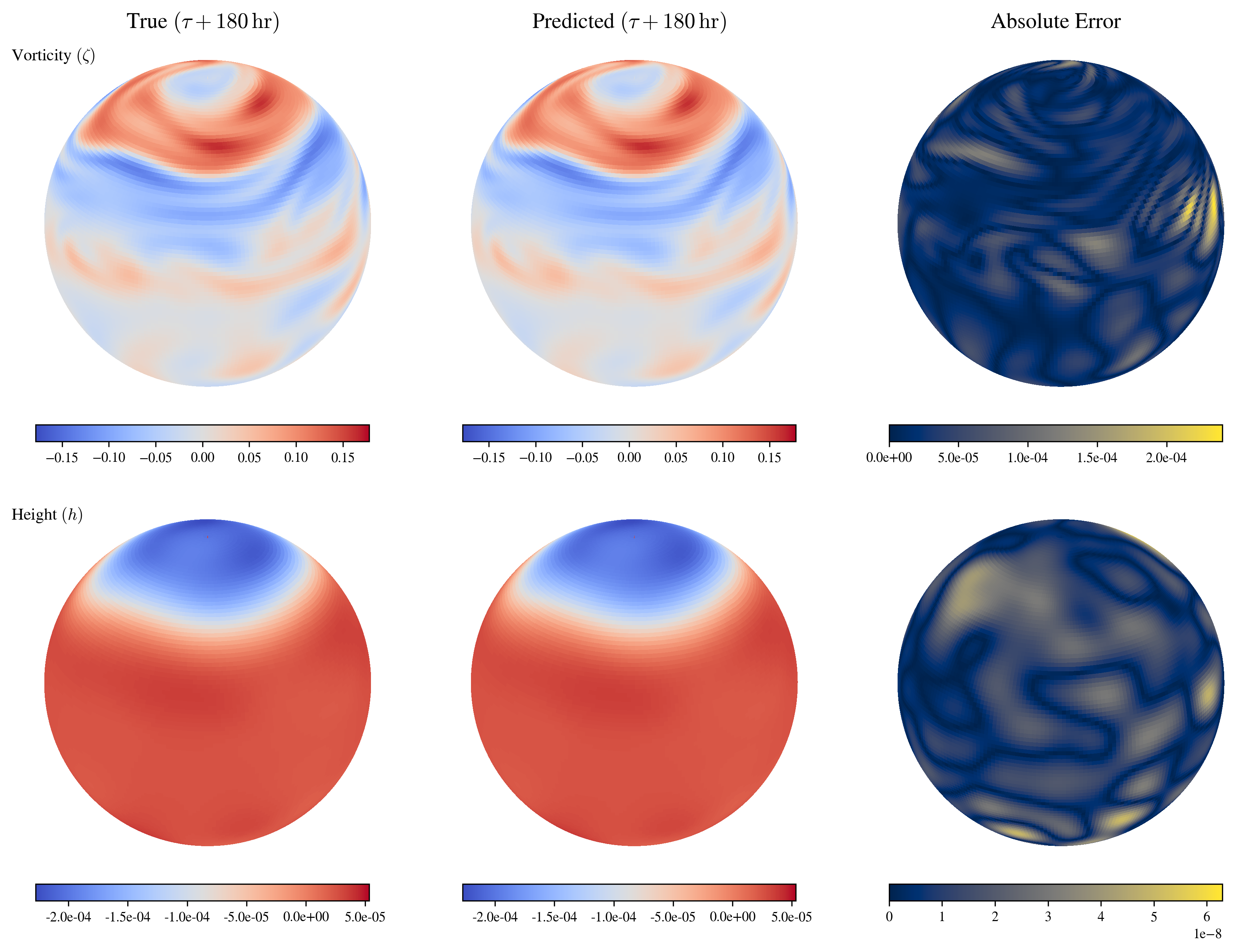}
    \vspace{5mm}
    \\
    \includegraphics[width=\linewidth]{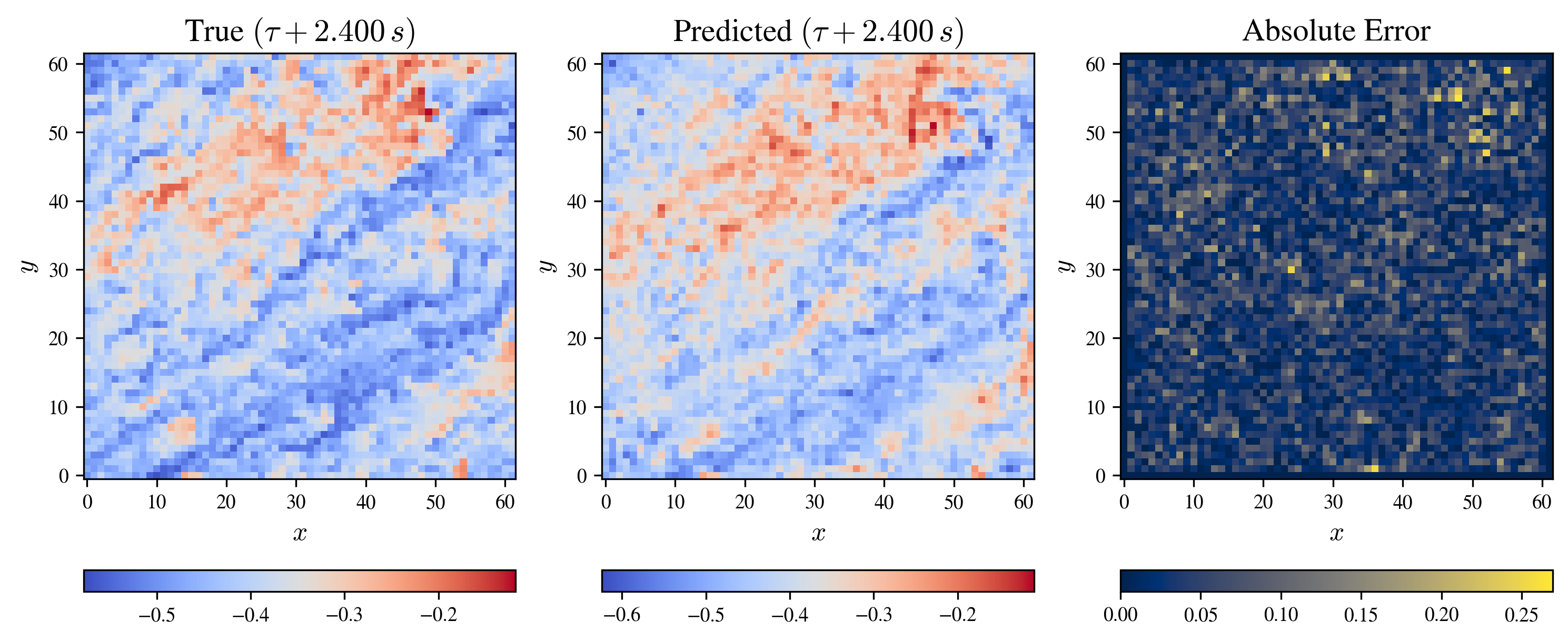}
    \caption{{\color{blue}Illustrating the forecast error from the discovered WSINDy model at the final time for the `Spherical' (top) and \lq{Stratified}' (bottom) datasets. See \S\ref{forecasting} for a discussion.}}
    \label{fig:spherical_forecast}
\end{figure}

\begin{figure}
    \centering
    \includegraphics[width=0.95\linewidth]{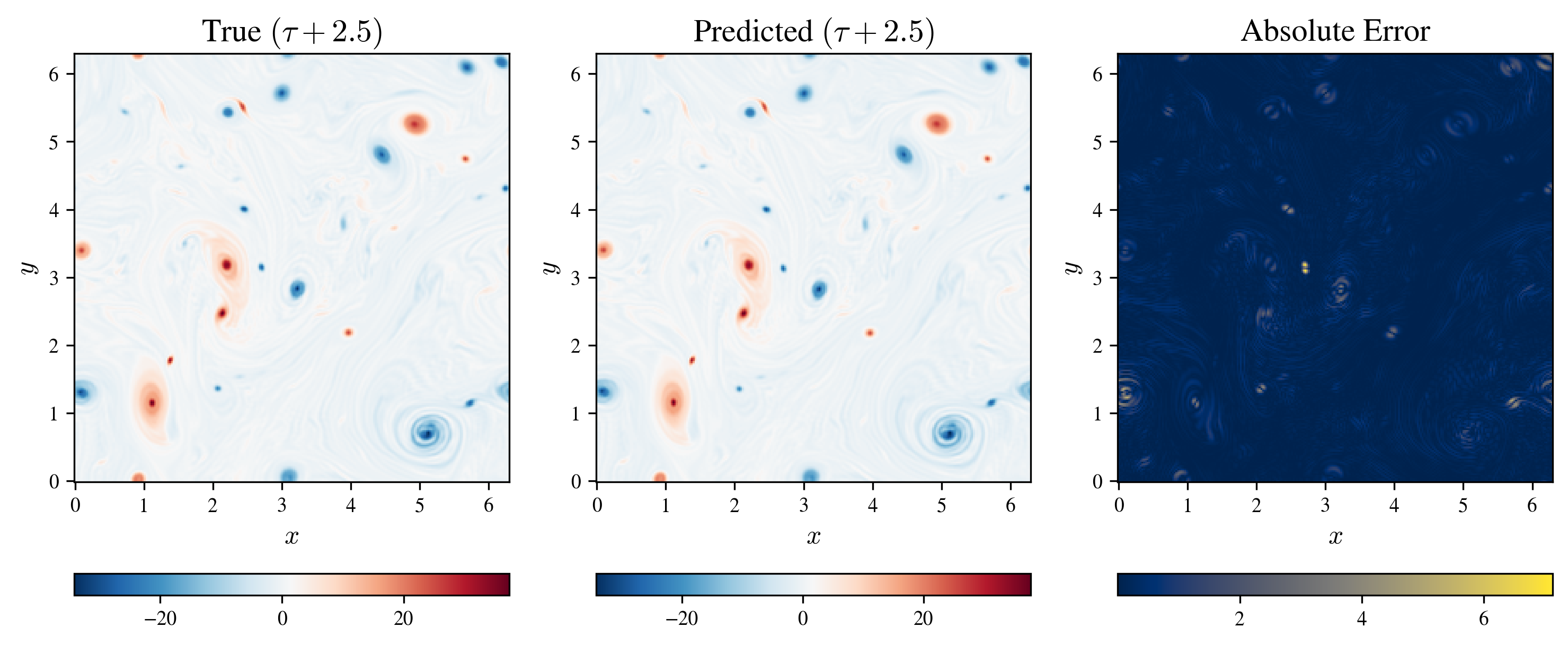}
    \includegraphics[width=0.95\linewidth]{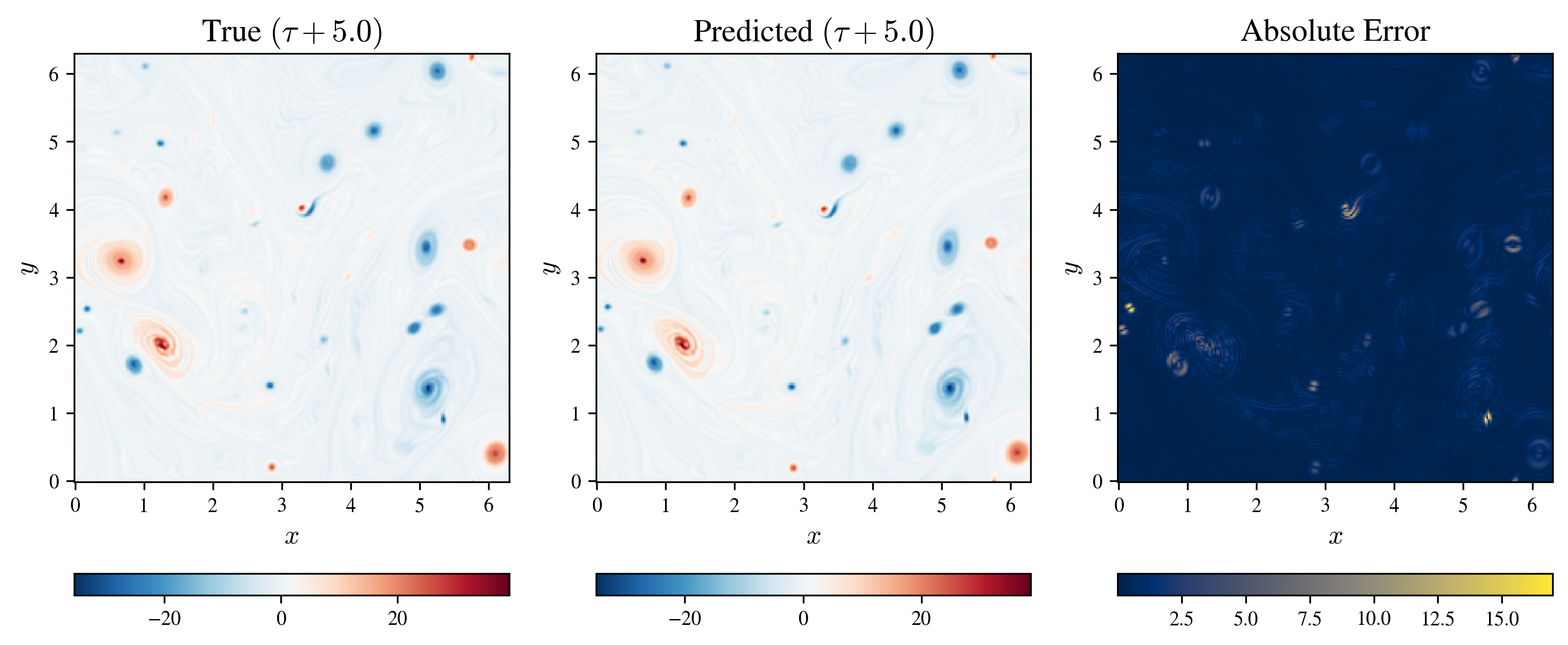}
    \includegraphics[width=0.95\linewidth]{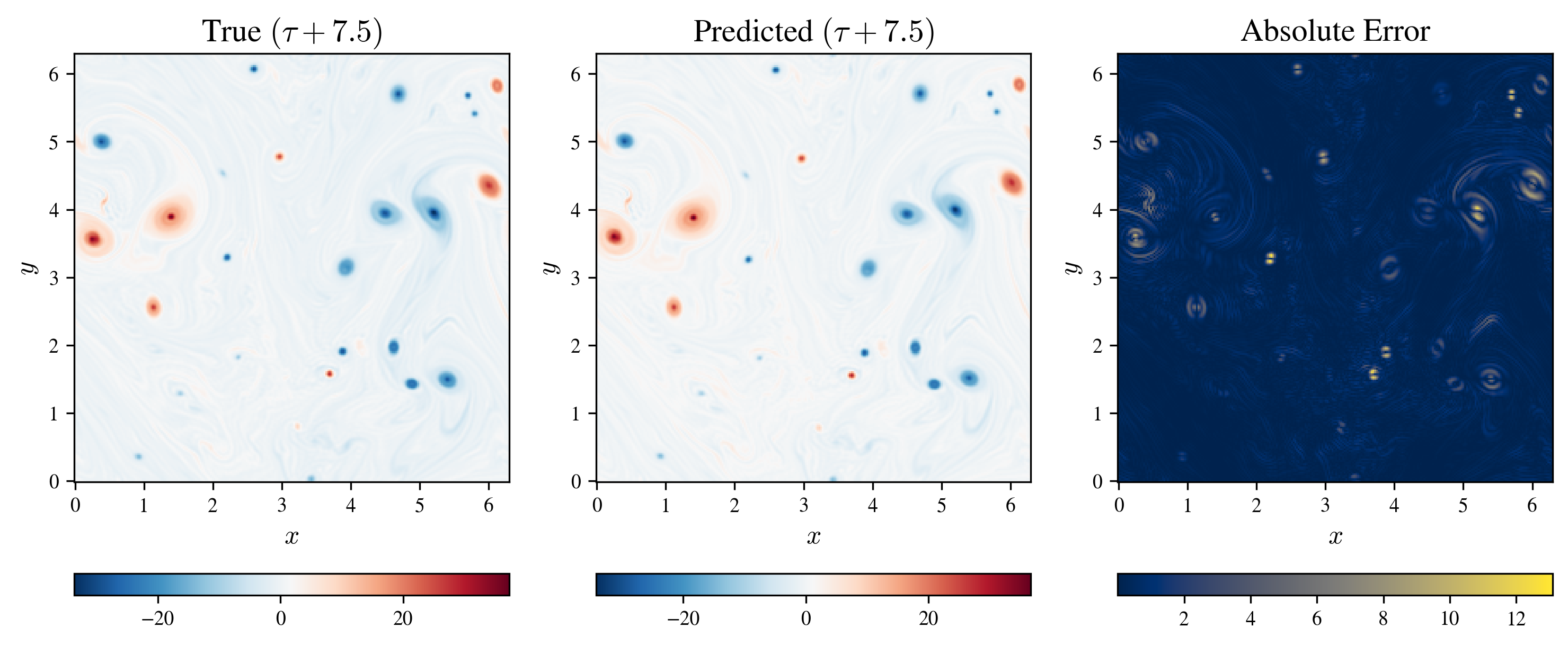}
    \includegraphics[width=0.95\linewidth]{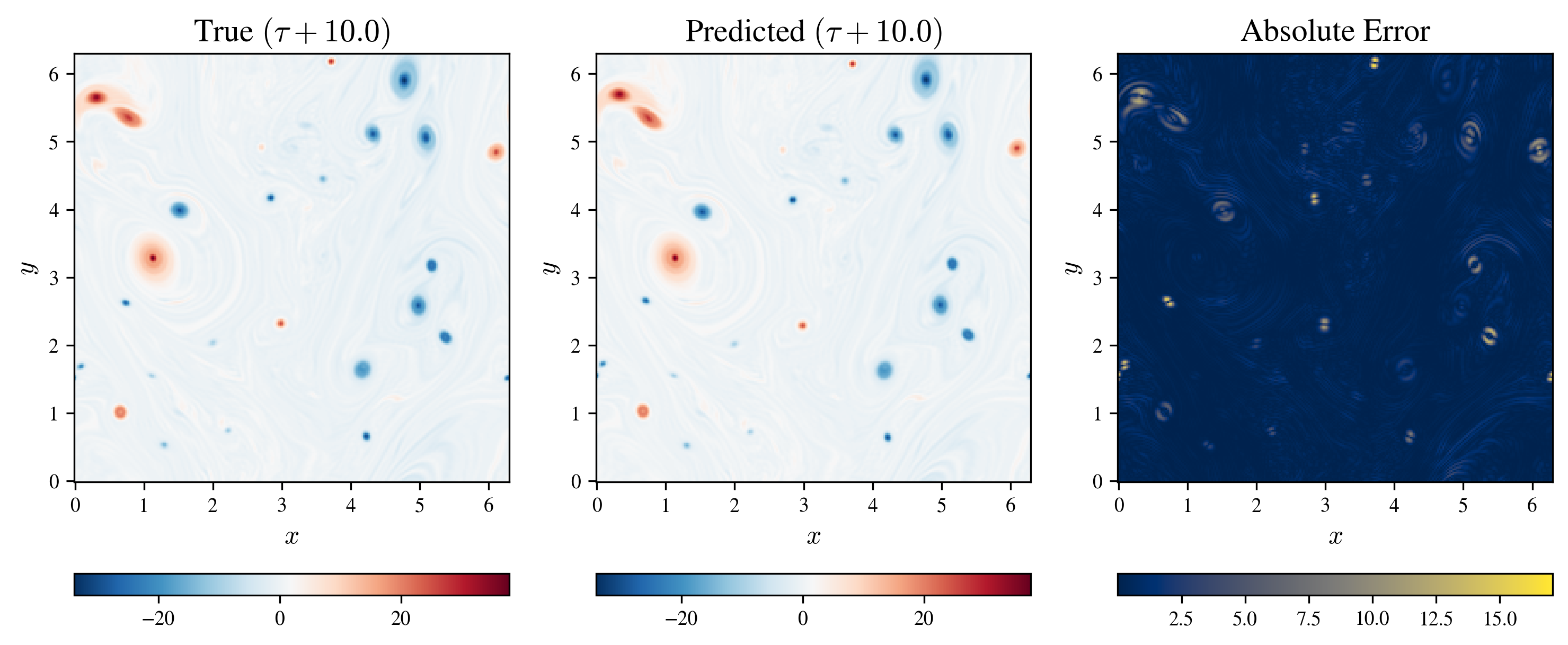}
    \caption{Rolling out a forecast for the `Barotropic' dataset for times $t \in [\tau, \tau+10]$.}
    \label{fig:barotropic_forecast}
\end{figure}

A striking advantage of symbolic model identification is the ability to produce accurate forecasts whenever the true form of the physics is recovered, which holds even in cases of highly turbulent data; see Figures~\ref{fig:spherical_forecast} and \ref{fig:barotropic_forecast}. This phenomena is exemplified by the small forecasting errors reported in Table~\ref{table:forecastingTable}, which demonstrate that a turbulent fluid state can be estimated in a point-by-point manner with a relative RMS error of $21\%$ {\color{red}or less} for test data extended half the duration of the training data past the final time {\color{blue}$\tau$}. {\color{red}As the} data {\color{red}become} less dynamic on the time scales of interest{\color{red}, quantified by increasing $T_0$ values (see Figure~\ref{fig:integral_timescales})}, the same error is observed shrink as low as $\mathcal{O}(10^{-2})$. {\color{red} Note that although the `Stratified' forecast in Table~\ref{table:forecastingTable} exhibits respectable average error metrics (i.e., RMSE and $\overline{\mathcal{E}}_F$ values), the poor $R^2 < 0.5$ indicates that the forecast $\mathbf{U}^+$ fails to account for more than half of the variance of the validation data $\mathbf{U}^{\star}$. In this case, we suspect that the low $R^2$ value is largely a reflection of the non-smooth nature of the underlying data (see Figure~\ref{fig:spherical_forecast}), rather than being due to model misspecification error.}

{\color{red}To give a crude idea of the state-of-the-art in NWP, \citeA{RaspHoyerMeroseEtAl2024JAdvModelEarthSyst} reported that at an RMS error of $\sim\!{2}$ m/$s$ given two days of lead time ($\sim\!2.7$\% better than the corresponding IFS forecast), GraphCast outperformed other popular models when predicting wind velocities at $p = 850$ hPa. Since wind speeds of $\sim\!15-25$ m/$s$ are typical at this pressure level, this result very roughly indicates an $\mathcal{O}(10)$ percent forecasting error. Although the analogy between the two cases is strained and we do expect GraphCast to outperform WSINDy (at least, in its current incarnation) in most realistic weather prediction tasks, the similar order of error listed in Table~\ref{table:forecastingTable} lends credibility to the notion that sparse regression approaches may eventually be competitive in this space.}


\section{Discussion} \label{conclusion}
We have detailed the application of the WSINDy algorithm to atmospheric fluid data, demonstrating that the approach is capable of identifying interpretable PDE models in several examples of geophysical interest. In particular, WSINDy recovers accurate governing equations for {\color{red}numerical} data drawn from turbulent fluid {\color{red}simulations} in both Euclidean and spherical domains. {\color{red}Moreover, WSINDy is capable of extracting latent meteorological relationships (such as effective conservation laws) from assimilated data, although the recovery of accurate and sparse momentum equations remains a challenge.} Since the governing model is learned in a symbolic form, physical phenomena governed by canonical PDEs such as the shallow water equation and the barotropic vorticity equation can be identified directly. Moreover, the weak form representation of the data allows for robust model discovery in the presence of observational noise. We have primarily aimed to demonstrate to the geophysics community that WSINDy represents a powerful tool for {\color{red}interpretable}, data-driven {\color{red}atmospheric} modeling.


{\color{blue}Philosophically, SINDy-based modeling approaches aim to discover an evolution equation that is consistent with one or several sequences of spatiotemporal data, which are in turn assumed to represent a family of solutions to the unknown PDE in question. Ideally, the symbolic nature of the discovered model is then not just useful in a predictive capacity but can also be understood analytically, potentially yielding new physical insights about the problem at hand. Alternatively, if an approximate or idealized governing equation is already known, these approaches can also be used to identify a} \lq{closure} model' {\color{blue}that increases the physical realism of the model. Because such approaches are fundamentally data-driven, it is important to keep in mind that the discovered models are} \lq{effective}' {\color{blue}in the sense that they represent the portion of the phase space sampled by the given trajectory or trajectories, but are not necessarily faithful to the true dynamics in regimes where data is not available; see \cite{VaseyMessengerBortzEtAl2025JComputPhys} for an in-depth exploration of this phenomenon in the context of plasmas. We also note that there has been recent work aimed at identifying structures such as attractors in the phase space of finite-dimensional systems; see, e.g.,} \cite{LemusHerrmann2024NonlinearDyn}.

{\color{red}In its current conception, the WSINDy paradigm does face several challenges and limitations. A salient example in our work is the relatively poor performance of WSINDy on the `ERA5' example vis-\`a-vis effective momentum equation discovery, which we see as its largest outstanding challenge for competitive forecasting performance against black-box models such as GraphCast \cite{LamSanchez-GonzalezWillsonEtAl2023Science} in weather and climate contexts. Although WSINDy was built to emphasize interpretability over expressivity, the encouraging results shown in Table~\ref{table:forecastingTable} indicate that if this barrier can be overcome, perhaps by innovations in the treatment of non-autonomous candidate terms or scale-dependent phenomena, sparse regression may also be an exciting tool in the sphere of forecasting.}

We conclude with a brief list of natural extensions of this work. In particular, an interesting (and not fully-understood) aspect of using a weak form representation of the dynamics is the ability to investigate a particular range of length and time scales by appropriately localizing the corresponding test functions. We suspect that this is a fruitful direction for future research, potentially bearing upon questions of scale dependence in the setting of data-driven modeling. For example, such work could serve as a foundation for a numerical framework capable of smoothly transitioning between weather and climate modeling as a function of the test function support radii $\boldsymbol{\ell}$. {\color{blue}Moreover, there are potentially interesting applications of weak form model identification in the setting of data assimilation. For example, WSINDy could be used to: (1) identify a data-driven \textit{assimilation model} from empirical observations, to later be used to produce assimilated data, or to (2) learn an accurate governing equation consistent with the assimilated data and compare this with the model used in the fit, quantifying the term-wise error and any physical mechanisms left unrepresented. While a more detailed statement about the potential for using the SINDy paradigm in tandem with data assimilation methods lies well beyond the scope of this paper, we suspect that the such work could result in algorithmic strategies for improving assimilation models.}

\section*{Open Research}
Our code is publicly available at: {\color{red}\url{https://github.com/MathBioCU/WSINDy4Atmos}}.

\section*{Acknowledgments}
The authors would like to thank Ian Grooms (University of Colorado) for guidance regarding weather models and data assimilation. The authors would also like to thank April Tran for insights regarding software development for weak form methods{\color{red}, as well as the Johns Hopkins Turbulence Database team for assistance with their database.}

This work is supported in part by National Science Foundation grants 2054085 and 2109774, 
National Institute of Food and Agriculture grant 2019-67014-29919, and Department of Energy grant DE-SC0023346.

\section*{Conflict of Interest Statement}
The authors have no conflicts of interest to disclose.


\bibliography{WSINDy4Atmos}

\newpage
\appendix

\RaggedRight

\section{Description of Datasets}\label{dataset_descriptions}
Our results encompass three distinct numerical simulations and one assimilated dataset, all of which are publicly available upon request of the referenced sources. {\color{blue}In what follows we detail these datasets, listing numerically relevant information in} Table~\ref{table:dataset_descriptions}.
\begin{enumerate}
    \item \lq\textbf{Spherical}': a 2D simulation of a mid-latitude unstable jet on the surface of a sphere, sourced from the Dedalus project \cite{BurnsVasilOishiEtAl2020PhysRevResearch};\\\url{https://dedalus-project.readthedocs.io/en/latest/pages/examples/ivp_sphere_shallow_water.html}
    \item \lq\textbf{Barotropic}': a 2D simulation of equivalent barotropic turbulence with doubly periodic boundary conditions, sourced from PyQG \cite{AbernatheyRochanotesRossEtAl2022};\\\url{https://pyqg.readthedocs.io/en/latest/examples/barotropic.html}
    \item \lq\textbf{Stratified}': a {3D LES of a stably-stratified atmospheric boundary layer} in a periodic cube, sourced from the Johns Hopkins Turbulence Database \cite{LiPerlmanWanEtAl2008JournalofTurbulence};\\\url{https://turbulence.idies.jhu.edu/datasets/geophysicalTurbulence/sabl} 
    \item \lq\textbf{ERA5}': {\color{blue}assimilated global weather data on a single pressure level ($200$ hPa) from the ECMWF Reanalysis v5} \cite{CopernicusClimateChangeService2023};\\\url{https://cds.climate.copernicus.eu/datasets/reanalysis-era5-pressure-levels}
\end{enumerate}
\begin{table}[htb]
\begin{center}
\begin{tabular}{||c c c c c c||} 
\hline
\textbf{Dataset} & \textbf{Grid Size} & \textbf{Observables} & \textbf{Terms} & \textbf{Query Pts.} & \textbf{Time ($\boldsymbol{s}$)} \\ [0.5ex] 
\hline\hline
Spherical & $256\!\times\!112\!\times\!360$ & See \S\ref{spherical} & $61$ & $3780$ & 137
\\ [1ex]
Barotropic & $256\!\times\!256\!\times\!160$ & $\{\zeta, u, v\}$ & $41$ & $77440$ & 77
\\ [1ex]
Stratified & $62\!\times\!62\!\times\!67$ & $\{\vartheta, u, v\}$ & $45$ & $62160$ & $< 1$
\\ [1ex]
ERA5 & $720\!\times\!341\!\times\!96$ & See \S\ref{era5_description} & $29$ & $6864$ & 126
\\ [1ex]
\hline
\end{tabular}
\end{center}
\caption{\color{blue} Numerical details about the datasets used in this paper. The uniform grid sizes are listed in $\Delta{x_1} \times \dots \times \Delta{x_n} \times \Delta{t}$ order, where we set $x_1 = \varphi$ and $x_2 = \theta$ for spherically gridded 2D data. The \lq{Observables}' column lists the variables $\{u_1, \dots, u_d\}$ that are considered accessible to the weak library $\mathbf{G}$ for the $\pt_t \mathbf{u}_1$ model. The \lq{Time}' column lists the total time in seconds required to build $\mathbf{G}$ on a 2-core Intel Xeon 2.2GHz CPU with 13 GB of RAM.}
\label{table:dataset_descriptions}
\end{table}

\newpage

\begin{figure}
    \centering
    \includegraphics[width=0.95\linewidth]{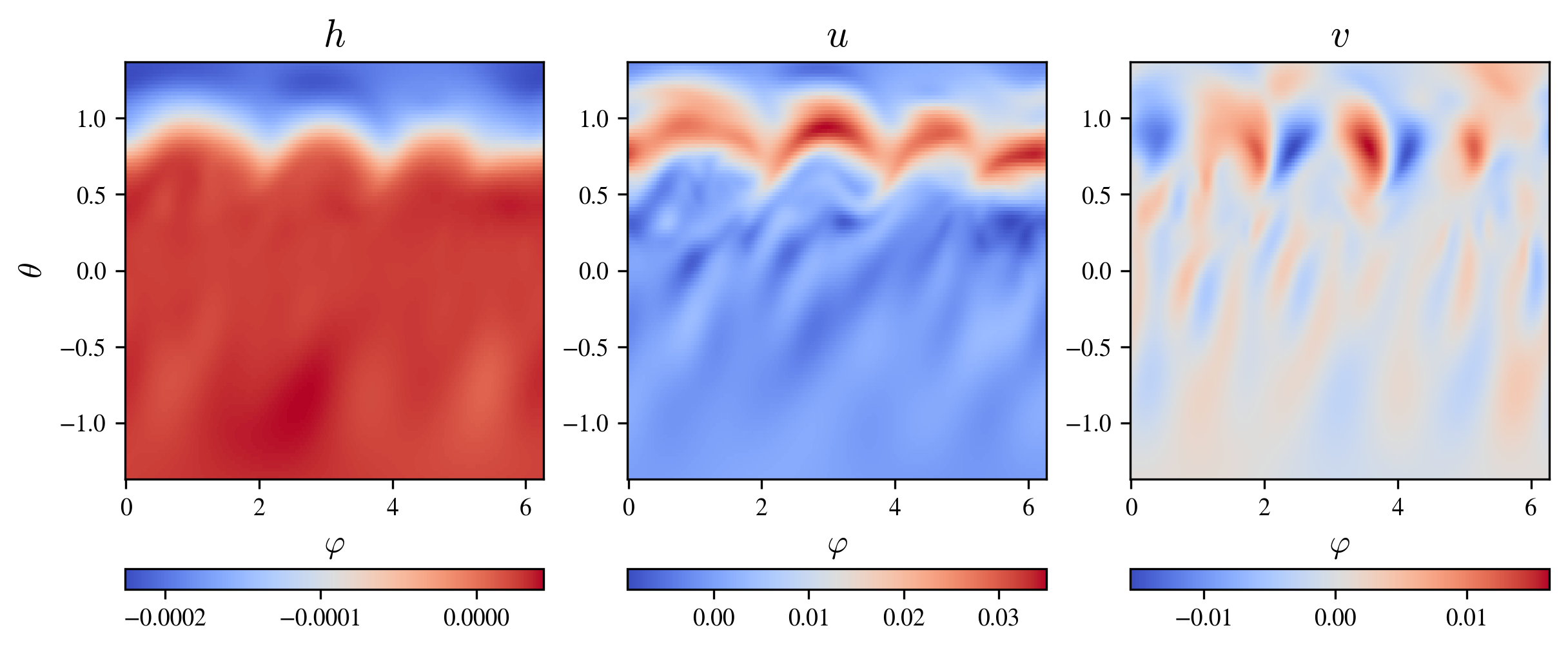}
    \caption{{\color{blue}A snapshot from the \lq{Spherical}' simulation of \S\ref{spherical}, created using Dedalus code \cite{BurnsVasilOishiEtAl2020PhysRevResearch} based upon the numerical benchmark proposed by \citeA{GalewskyScottPolvani2004TellusDynMeteorolOceanogr}.}}
    \label{fig:spherical_appendix_plot}
\end{figure}

\subsection{Shallow Water Equations on a Sphere (`Spherical')}\label{spherical}
This simulation, proposed as a numerical benchmark by  \citeA{GalewskyScottPolvani2004TellusDynMeteorolOceanogr}, models the evolution of an unstable mid-latitude tropospheric jet undulating under the influence of a small initial perturbation {\color{blue}(see Figure~\ref{fig:spherical_appendix_plot})}. The underlying mathematical model is the viscous shallow-water equation, solved on the surface of a sphere. Using rescaled units in which the radius $r = 1$ and $\Omega = 0.263$, the equations governing the free surface of the fluid layer $h = h(\varphi,\theta)$ and the large-scale geophysical flow $\boldsymbol{u} = \boldsymbol{u}(\varphi, \theta)$ are
\begin{align*}
\begin{cases}
h_t + \nabla\cdot(h\boldsymbol{u}) \, = - H_0(\nabla\cdot\boldsymbol{u}) + \nu\Delta^2h, \\
\boldsymbol{u}_t + (\boldsymbol{u}\cdot\nabla)\boldsymbol{u} = \boldsymbol{f} - g\nabla{h} + \nu\Delta^2\boldsymbol{u},
\end{cases}
\quad \text{with parameters} \quad
\begin{cases}
H_0 = 1.57\cdot{10^{-3}}, \\
g = 19.947,
\end{cases}
\end{align*} where $\nu\Delta^2$ is the hyperviscosity operator with $\nu = 8.66\cdot{10^{-9}}$. Since the hyperviscosity terms are used for numerical stability, we omit these terms from the vector of true coefficients $\mathbf{w}^{\rm{true}}$ and include an asterisk next to the corresponding results in Table~\ref{table:simulated}.

A benefit of the Dedalus framework is that it is straightforward to write out direct observations of the transport quantities (e.g., $\mathfrak{A}$ and $\mathfrak{D}$) during the simulation process. 
Due to the cumbersome form that these operators take in spherical coordinates, we explicitly include these observations in the {\color{red}set $\mathcal{U}$ of} state vector measurements, using {\\}${\color{red}\mathcal{U} = } \{h, \boldsymbol{u}, \nabla\cdot(h\boldsymbol{u}), \nabla\cdot\boldsymbol{u}, \nu\Delta^2h \}$ and ${\color{red}\mathcal{U} =} \{ \boldsymbol{u}, (\boldsymbol{u} \cdot \nabla)\boldsymbol{u}, \boldsymbol{f}, \nabla{h}, \nu\Delta^2\boldsymbol{u} \}$ for the $h_t$ and $\boldsymbol{u}_t$ models, respectively. We include $\{\mathcal{D}^if_j\}$ terms evaluated on $h, \, u,$ and $v$ in the weak library $\mathbf{G}$ as per eq.~(\ref{eq:standardizedLibrary}), which is computed over $3780$ query points and contains $61$ terms for the $h_t$ model and $41$ terms for the $u_t$ and $v_t$ models. In each case, we use support parameters $\boldsymbol{\ell} = (30, \, 14, \, 35)$. To avoid numerical issues caused by singularities at the poles $\theta = \pm \pi/2$, we trim the data to latitudes of roughly $\theta \in [-89.8^{\circ}, 89.8^{\circ}]$.


\begin{figure}
    \centering
    \includegraphics[width=1\linewidth]{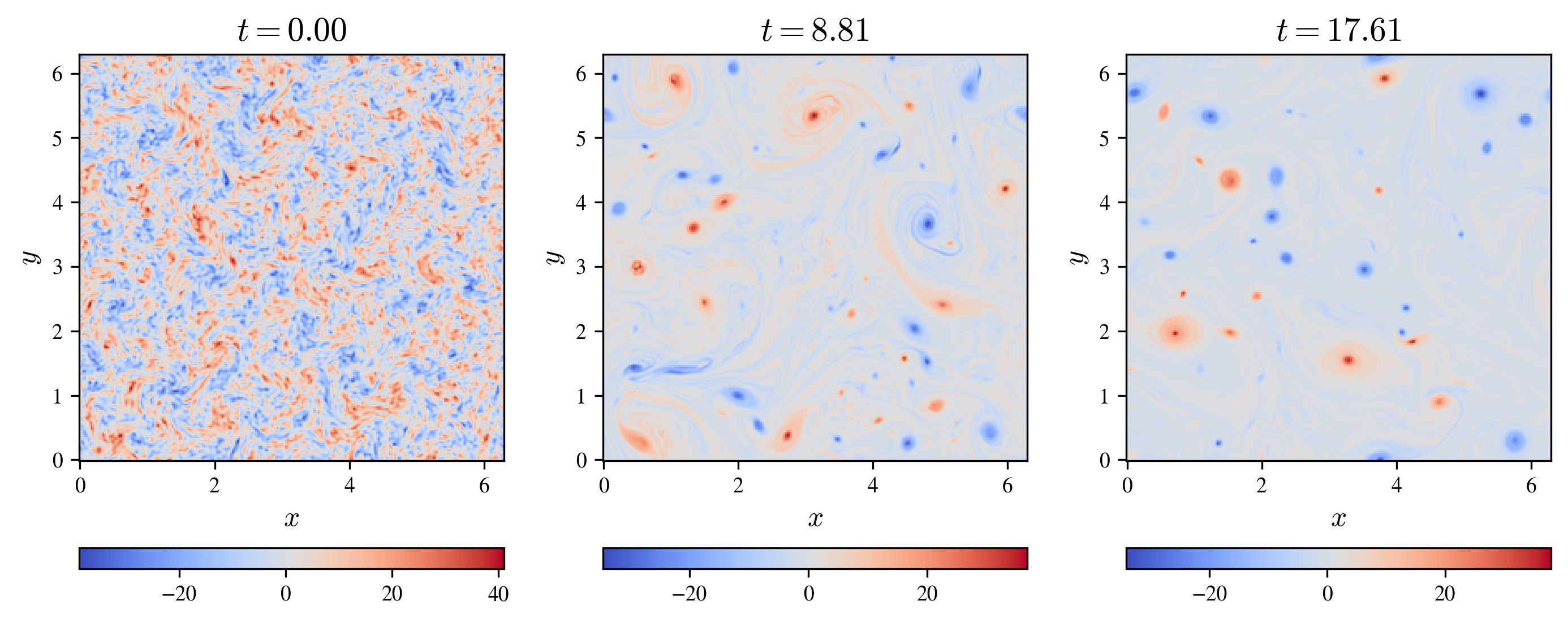}
    \caption{Snapshots from the numerical simulation of equivalent barotropic turbulence in \S\ref{barotropic}, made using PyQG code \cite[]{AbernatheyRochanotesRossEtAl2022} based upon the turbulence study by \citeA{Mcwilliams1984JFluidMech}. The scalar vorticity field ${\color{blue}\zeta = |\!| \nabla \times \boldsymbol{u} |\!|_2}$ is plotted.}
    \label{fig:pyqg_simulation}
\end{figure}

\subsection{Equivalent Barotropic Turbulence (`Barotropic')}\label{barotropic}
The equivalent barotropic simulation is an idealized model which represents incompressible horizontal wind flow in a regime without temperature gradients or misalignment of pressure and density gradients, which would normally give rise to more complex dynamics. This example uses a special initial condition based on a turbulence study by \citeA{Mcwilliams1984JFluidMech}, in which the vorticity field $\zeta = \zeta(x,y,t)$ is initialized according to a prescribed initial frequency spectrum $|\hat{\psi}_0| \propto (k[ 1 + (k/6)^4 ])^{-1}$ for the stream function $\psi$ (where $\zeta = \Delta\psi$). This initial condition is chosen because it gives rise to coherent vortex structures (see Figure~\ref{fig:pyqg_simulation}).

Numerically, this simulation uses a uniform grid spacing of $\Delta{x}=\Delta{y}=2\pi/256$ and is integrated for times $t\in[0,30]$. The evolution equation for the vorticity $\zeta$ is given by the advection equation, \begin{align}\label{eq:pyqg}
    \begin{cases}
        \zeta_t + (\boldsymbol{u} \cdot \nabla)\zeta = \texttt{ssd},
        \\
        \nabla\cdot\boldsymbol{u} = 0,
    \end{cases}
\end{align} where \texttt{ssd} represents the influence of a ``small-scale dissipation'' model, which is achieved by implementing a ``highly selective spectral exponential filter'' term during the integration process \cite{AbernatheyRochanotesRossEtAl2022}. Because this term is difficult to characterize and only affects large wavenumbers, we report results in Table~\ref{table:simulated} as though \texttt{ssd} $=0$ and include a dagger ($\dagger$) next to the result (cf. the hyperviscosity terms in \S\ref{spherical}).

We use observations of the scalar fields ${\color{red}\mathcal{U} =} \{\zeta, u, v\}$ within a 41-term library over 77440 query points. When discovering models for the horizontal divergence $\nabla\cdot\boldsymbol{u}$, we instead set $\mathcal{D}^0=\pt_x$ and use ${\color{red}\mathcal{U} =} \{u,v\}$ within a {\color{red}30}-term library {\color{red}(including $u_t$, $v_t$, etc.)}, discovering an equation of the form $u_x\approx-v_y$. We set the effective length and time scales by using support radii given by $\boldsymbol{\ell}=(20, \, 20, \, 20)$ and $\boldsymbol{\ell}=(38, \, 35, \, 20)$, respectively.


\begin{figure}
    \centering
    \includegraphics[width=\linewidth]{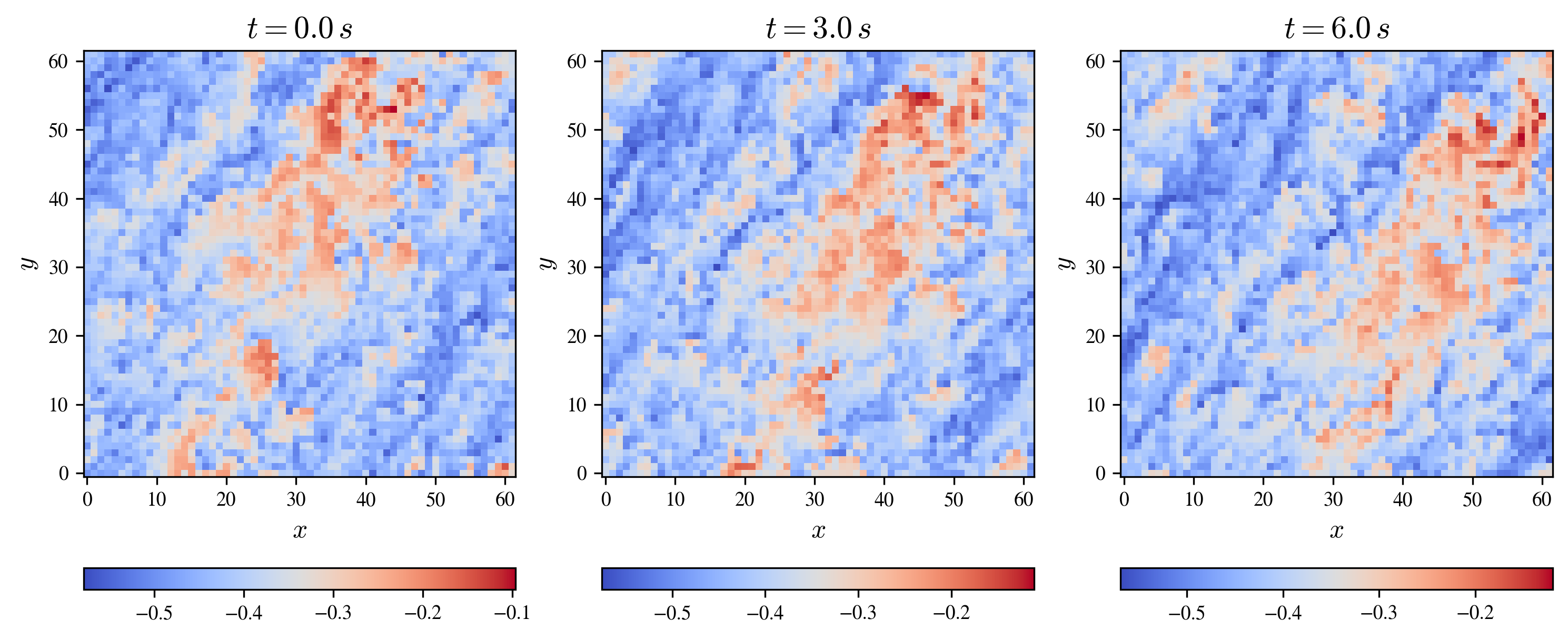}
    \caption{Snapshots of the potential temperature field $\vartheta$ from the \lq{Stratified}' dataset (\S\ref{stratified}) in the $(x,y)$ plane at a constant height $z = 0$, where warmer colors indicate higher temperatures. The flow is driven by a constant geostrophic wind $\boldsymbol{v}_g$ that causes traveling wave dynamics; the corresponding Weak SINDy model is a effective transport equation (see Table~\ref{table:simulated}). 
    }\label{fig:geophysical_wave}
\end{figure}

\subsection{Stably-Stratified Atmospheric Boundary Layer (`Stratified')}\label{stratified}
The final simulated dataset, modeling a stably-stratified atmospheric boundary layer, is sourced from a large-eddy simulation (LES) of the incompressible Boussinesq equations governing the wind velocity $\boldsymbol{v} = \boldsymbol{v}(x,y,z,t)$ and potential temperature $\vartheta = \vartheta(x,y,z,t)$ inside of a periodic cube $\mathcal{X} = [0, \, 400]^3$ (meters) for times $t \in [0, {\color{red}7.5}]$ (seconds). Readers interested in the numerical details are directed towards the recent original work by \citeA{McWilliamsMeneveauPattonEtAl2023Atmosphere}. Here, we investigate a heavily sub-sampled 2D slice of the spatial domain $\mathcal{X}$ at a constant height $z_0 = 0$, with $(x,y) \in [0, \, 62]^2$ subject to a uniform discretization of $\Delta x = \Delta y = 1$ (meter). The \textit{spatially filtered} potential temperature $\vartheta = \mathbf{F}[\vartheta']$ (here, $\mathbf{F}$ denotes the LES filter) is governed by \begin{align}\label{eq:stratified_temp_model}
    \vartheta_t = - (\boldsymbol{v} \cdot \nabla) \vartheta - \nabla\cdot\mathbf{B},
    \quad \text{where} \quad 
    \mathbf{B} := \mathbf{F}[\vartheta' \boldsymbol{v}'] - \vartheta\boldsymbol{v},
\end{align} with $\boldsymbol{v} = \mathbf{F}[\boldsymbol{v}']$ denoting the filtered wind velocity, which is dominated by a constant geostrophic wind $\boldsymbol{v}_g$. In our case, the matrix $\mathbf{B}$ in eq.~(\ref{eq:stratified_temp_model}), which represents a subfilter-scale temperature model, cannot be explicitly constructed since neither the unfiltered temperature $\vartheta'$ nor velocity $\boldsymbol{v}'$ data are available. As a consequence, we do not list a $E_{\infty}$ coefficient error in the corresponding WSINDy results. We use {\color{red}observations} ${\color{red}\mathcal{U} =} \{\vartheta, u, v\}$ within a 45-term library computed over 62160 query points to discover an effective temperature model. For this example, we set $\boldsymbol{\ell} = (11, \, 10, \, 15)$.

\begin{figure}
    \centering
    \includegraphics[width=0.9\linewidth]{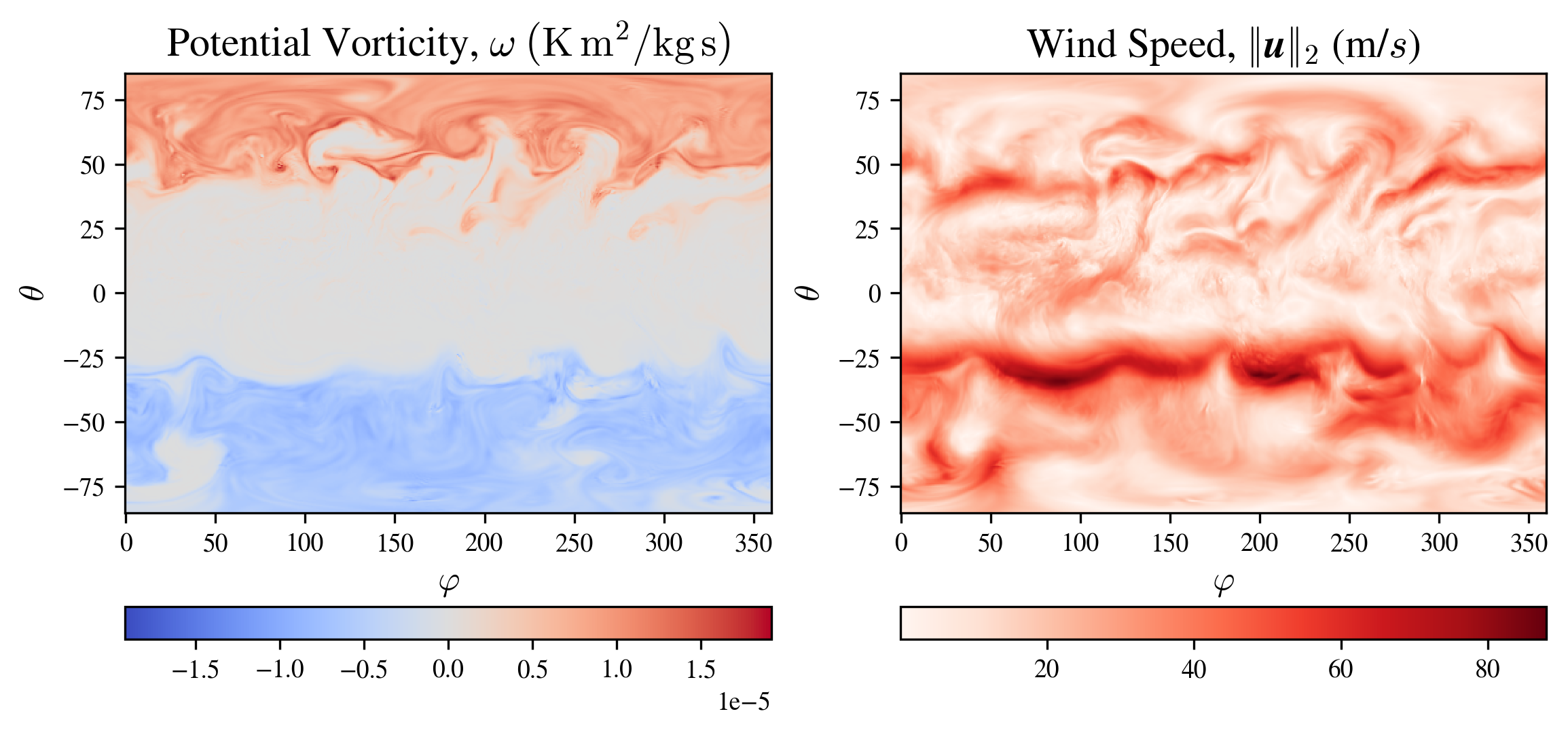}
    \caption{A snapshot of global ERA5 data \cite{CopernicusClimateChangeService2023} on a single pressure level ($p=200$ hPa). We use data in the latitudes $\theta \in [-85^{\circ}, 85^{\circ}]$.}
    \label{fig:era5data}
\end{figure}

\subsection{ERA5 Weather Data (`ERA5')}\label{era5_description}
These data represent hourly snapshots {\color{red}of various atmospheric variables} taken at a single pressure level of $200$ hPa (roughly corresponding to the upper Troposphere) in a global-scale domain $(\varphi,\theta) \in [0^{\circ}, 360^{\circ}] \times [-85^{\circ}, 85^{\circ}]$, beginning 00:00 UTC on July $25^{\rm{th}}$, 2024 and ending 23:00 UTC on July $28^{\rm{th}}$, 2024. We coarsen the original data into longitudinal and latitudinal resolutions of $\Delta\varphi, \, \Delta\theta = 1.25^{\circ}$ and approximate the altitude as $r \approx a = 6371$ km, using the corresponding {\color{red}assimilated observations} of the horizontal velocity $\boldsymbol{u}$, horizontal divergence $(\nabla \cdot \boldsymbol{u})$, vertical velocity $\dot{\eta}$, geopotential $\Phi$, and potential vorticity $\omega$. Here, the vertical wind velocity $\dot{\eta}$ is expressed in terms of the pressure-based coordinate $\eta$ reported in \cite{ECMWF2021IFSDocumentationCY47R3}.

To improve the condition number of the weak library, $\kappa(\mathbf{G})$, we manually rescale the potential vorticity data via $\omega \mapsto \omega/\Omega$. In the model discovery process, we use observations of the following set of {\color{red}atmospheric} variables: \begin{align*}
    {\color{red}\mathcal{U} =} \left\{\omega, \, \frac{\tan(\theta)\omega}{a}, \, \frac{u}{a\cos(\theta)}, \, v, \, w, \, \nabla\cdot\boldsymbol{u} \right\},
\end{align*} for the $\omega_t$ model, {\color{red}and} \begin{align*}
    {\color{red}\mathcal{U} =} \left\{u, \, \frac{u}{a\cos(\theta)}, \, \frac{\tan(\theta)u}{a}, \, v, \, w, \, \Phi, \, \nabla\cdot\boldsymbol{u} \right\},
\end{align*} for the $u_t$ and $v_t$ models. When discovering an evolution equation for the potential vorticity $\omega$, we set $\mathcal{D}^i \in \{1, \pt_x, \pt_y\}$ (also including $\mathfrak{D}(\omega)$ in the library) and use candidate terms $\mathcal{D}^i f_j(\varphi, \theta, \omega, u, v, w)$, with \begin{align*}
    f_j \in \left\{ 1, \, \omega, \, \frac{u}{a\cos(\theta)}, \, \frac{\omega u}{a\cos(\theta)}, \, v, \, \omega{v}, \, \frac{\tan(\theta)\omega{v}}{a}, \, w, \, \omega{w}, \, (\nabla \cdot \boldsymbol{u})\omega \right\}.
\end{align*} Analogously, when discovering a model for the wind velocity $\boldsymbol{u}$, we use candidate functions given by \begin{align*}
    f_j \in \left\{ 1, \, u, \, \frac{u}{a\cos(\theta)}, \, \frac{u^2}{a\cos(\theta)}, \, v, \, w, \, uv, \, \frac{\tan(\theta){uv}}{a}, \, uw, \, (\nabla \cdot \boldsymbol{u})u, \, (\nabla \cdot \boldsymbol{u})v, \, \Phi \right\}.
\end{align*} Here, we compute the results over 4576 query points and use test function support radii of $\boldsymbol{\ell} = (25, \, 25, \, 9)$. Physically, this choice of $\boldsymbol{\ell}$ corresponds to {\color{red}\lq{synoptic}'} length scales on the order of $30^{\circ}$ and a temporal scale of roughly $10$ hours{\color{red}, which are representative of the scales used in some general circulation models. Note that large-scale coherent weather patterns are expected to be relevant at the chosen pressure level of $p = 200$ hPa, such as the jet streams typically located nearby at $\sim250-300$ hPa.}

\section{IFS Primitive Equations}\label{IFS_eqns}
The IFS implements a primitive equation model for the horizontal wind velocity; see \cite{ECMWF2021IFSDocumentationCY47R3} for a detailed description. For reference, we list a simplified version of this model here: \begin{align}
    &u_t + (\boldsymbol{u} \cdot \nabla)u - \frac{\tan(\theta) uv}{a} + \dot{\eta} u_{\eta}
    - fv + \frac{\Phi_{\varphi}}{a\cos(\theta)} = \mathcal{R}_u, \label{eq:IFSmomentum_u}
    \\
    &v_t + (\boldsymbol{u} \cdot \nabla)v + \frac{\tan(\theta) u^2}{a} + \dot{\eta} v_{\eta}
    + fu + \frac{\Phi_{\theta}}{a} = \mathcal{R}_v,
    \label{eq:IFSmomentum_v}
\end{align} where $(\boldsymbol{u} \cdot \nabla)(\,\cdot\,) = \mathfrak{A}(\,\cdot\,)$ is the spherical advection operator of eq.~(\ref{eq:advection_operator}), with $a$ being the radius of the Earth, and $\eta$ is a hybrid vertical coordinate satisfying $\dot{\eta} \propto -w$ at a constant pressure levels $p$. Here, the $\mathcal{R}_u$ and $\mathcal{R}_v$ terms represent additional contributions to the dynamics due to horizontal diffusion and other parameterized physics.

\newpage

{\color{blue}
\section{Forecast Details}\label{sec:forecast_details}
The Dedalus \cite{BurnsVasilOishiEtAl2020PhysRevResearch} and PyQG \cite{AbernatheyRochanotesRossEtAl2022} frameworks used to simulate the \lq{Spherical}' (\S\ref{spherical}) and \lq{Barotropic}' (\S\ref{barotropic}) datasets, respectively, straightforwardly allow for forward-integration of the discovered WSINDy model in time. In these cases, we use built-in pseudo-spectral solvers from each framework to integrate the discovered model, using the final training data snapshot as an initial condition. Note that the corresponding numerical stability terms (i.e., the hyperviscosities and sub-grid dissipation model) are used in the simulated forecast of the discovered model. For the forecast of the \lq{Stratified}' dataset, we manually discretize the domain and integrate the discovered model in time using a standard $4^{\rm{th}}$ order Runge-Kutta scheme. Since the Stratified data is not smooth enough to rely upon finite-difference approximations for the spatial derivatives, we instead use a Savitzky-Golay derivative filter with a window size of 7 and a polynomial order of 3. {\color{blue} When reporting $t_{\rm{tol}}/T_0$ (see Figure~\ref{fig:integral_timescales}), we define the integral timescale as follows: \begin{align*}
            T_0 := {\color{red}\mathbb{E}_{ij}\left[\sigma_{ij}^{-2}\int_0^{T} \mathbb{E}_t\!\left[ u'_{ij}(t) \, u'_{ij}(t - \tau) \right] \, d\tau\right]}.
\end{align*} Here, $u'_{ij}(t) := u_{ij}(t) - \bar{u}_{ij}$ is the fluctuation for a single degree of freedom $u(x_i, y_j, t)$. In each case, we have used the longitudinal $u$-component of the velocity $\boldsymbol{u}$ to compute $T_0$ -- specifically, this is the main flow direction in the only case of anisotropic flow.}
}


\section{Scale-Invariant Preconditioning}\label{scale_invariance}
Consider a state vector $\boldsymbol{u} = [u_1, \dots, u_d] \in \mathbb{R}^d$, and suppose that the $l^{\rm{th}}$ component, $u_{l} = u_{l}(\boldsymbol{x},t)$, is a scalar field in $(n + 1)$-dimensions satisfying a PDE given by \begin{align}\label{eq:A1}
    \pt_t u_{l} = \sum_{i=1}^{I}\sum_{j=1}^{J} w(i,j) \, \mathcal{D}^i f_j(\boldsymbol{u}).
\end{align} Here, we assume each $f_j$ is a homogeneous function of degree $|\beta^j|$, given a corresponding multi-index $\beta^j = (\beta^j_1, \dots, \beta^j_d)$ for each of the fields $u_1, \dots, u_d$. In particular, we consider pairwise monomials of the form $f_j(\boldsymbol{u}) = \prod_{k=1}^{d} u_{k}^{\beta_{k}^j}$. Following \citeA{MessengerBortz2021JComputPhys}, we introduce a set of rescaled spatial and temporal coordinates via \begin{align*}
    (\tilde{x}_1, \ \dots, \ \tilde{x}_n, \ \tilde{t}) := (\gamma_{x_1}x_1, \ \dots, \ \gamma_{x_n}x_n, \ \gamma_tt),
\end{align*} where for each $\ell = 1, \dots, d$, \begin{align*}
    \tilde{u}_{l}\big(\tilde{\boldsymbol{x}}, \tilde{t}\big) :=
    \gamma_{u_{l}}u_{l}\left( \frac{\tilde{x}_1}{\gamma_{x_1}}, \dots, \frac{\tilde{x}_n}{\gamma_{x_n}}, \frac{\tilde{t}}{\gamma_{t}} \right)
    = \gamma_{u_{l}} u_{l}(\boldsymbol{x},t).
\end{align*}

If $u_{l}$ is a solution to eq.~(\ref{eq:A1}) above, then $\tilde{u}_{l}$ obeys the same PDE in the rescaled coordinates: \begin{align}\label{eq:A2}
    \tilde{\pt_t} \tilde{u}_{l}
    = \sum_{i=1}^{I}\sum_{j=1}^{J} \tilde{w}(i,j) \, \tilde{\mathcal{D}}^i f_j(\boldsymbol{\tilde{u}}).
\end{align} We relate the scaled and original weights ($\tilde{\mathbf{w}}$ and $\mathbf{w}$, respectively) via the change of coordinates $\mathbf{w} = \boldsymbol{\mu}^T \tilde{\mathbf{w}}$, where \begin{align*}
    \mu(i,j) := \frac{1}{\gamma_{u_{l}}} \left[ \prod_{k=1}^{d} \gamma_{u_k}^{\left(\beta^j_k\right)} \right]
    \left[\prod_{r=1}^{n} \gamma_{x_r}^{\left(\alpha^0_{r}-\alpha^i_{r}\right)} \right]\gamma_{t}^{\left(\alpha^0_{t}-\alpha^i_{t}\right)}.
\end{align*} {\color{blue}Here, each $\alpha^i = (\alpha_1^i, \dots, \alpha_n^i, \alpha_{n+1}^i)$ acts as a multi-index on the differential operator $\mathcal{D}^i = \mathcal{D}^{\alpha^i}$, with \begin{align*}
    \mathcal{D}^{\alpha^i} = \frac{\pt^{|\alpha^i|}}{\pt_{x_1}^{\alpha_1^i} \cdots \pt_{x_n}^{\alpha_n^i} \pt_t^{\alpha^i_{n+1}}}.
\end{align*}} The linear system in eq.~(\ref{eq:wsindy}) is then constructed in the rescaled coordinates of eq.~(\ref{eq:A2}), where we pick scaling factors $\gamma_u, \gamma_{x_i}, \gamma_t$ such that the condition number of the rescaled library is improved, with $\kappa(\tilde{\mathbf{G}}) < \kappa(\mathbf{G})$.

\begin{figure}[htb!]
    \centering
    \includegraphics[width=0.495\linewidth]{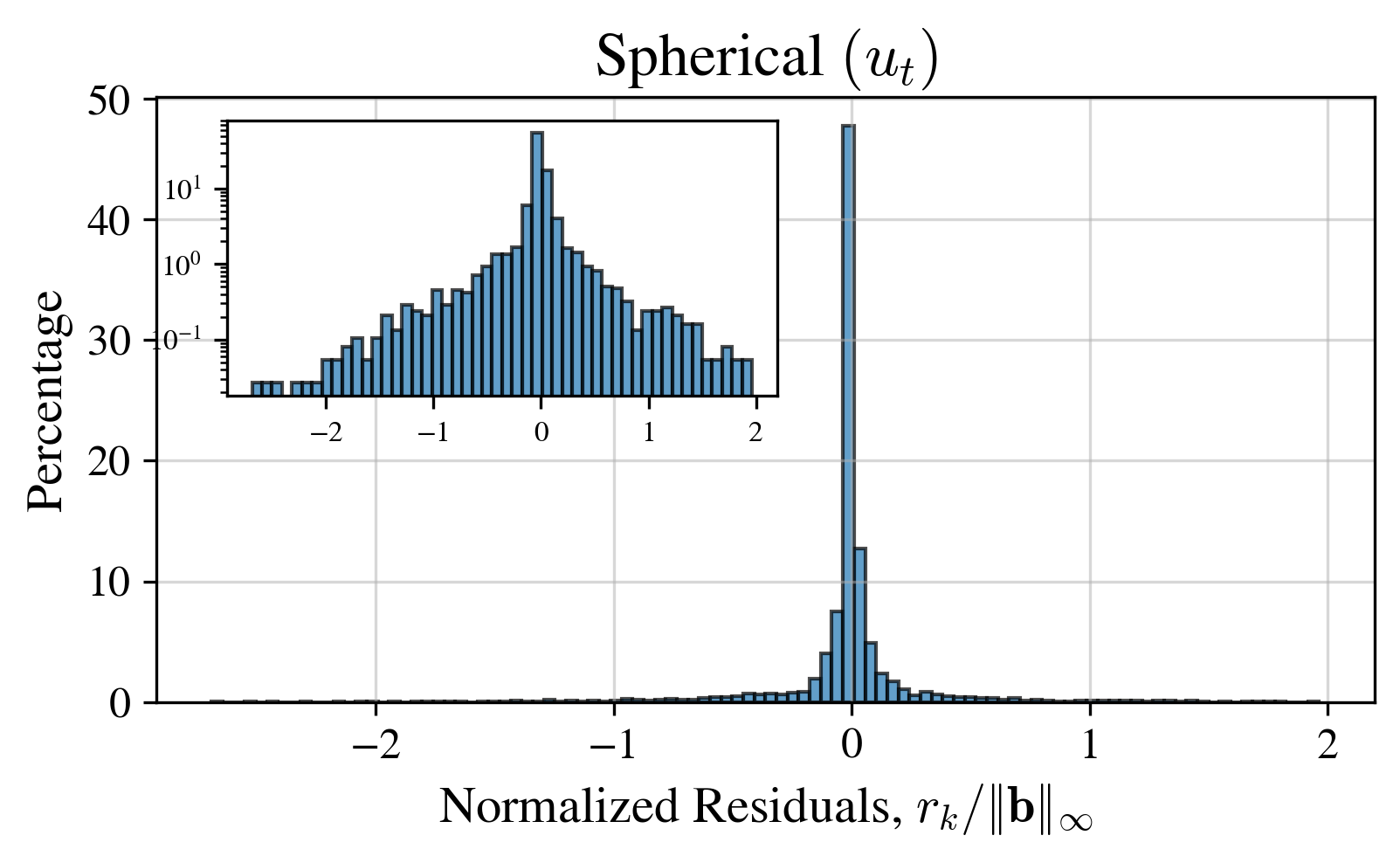}
    \includegraphics[width=0.495\linewidth]{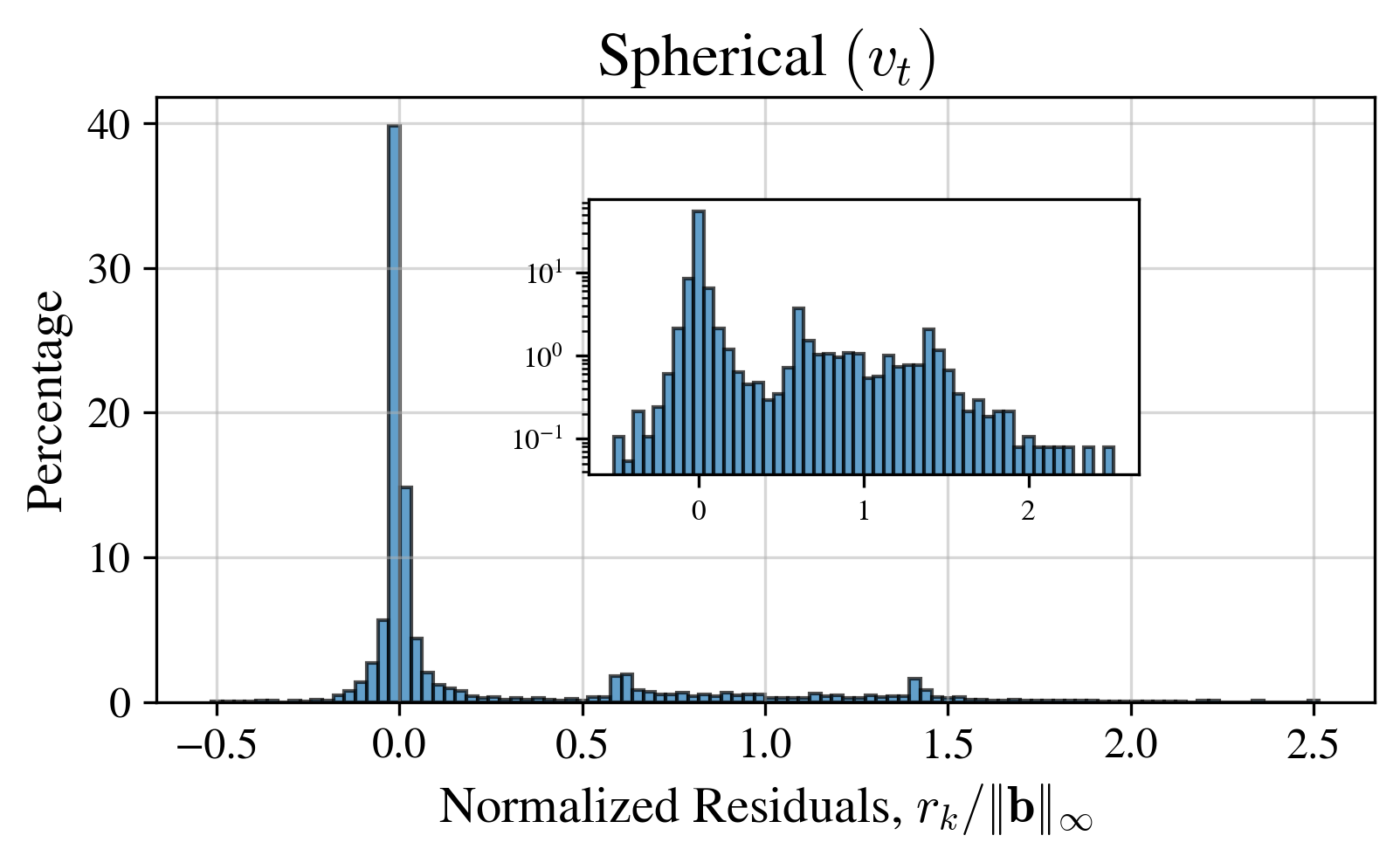}
    \includegraphics[width=0.485\linewidth]{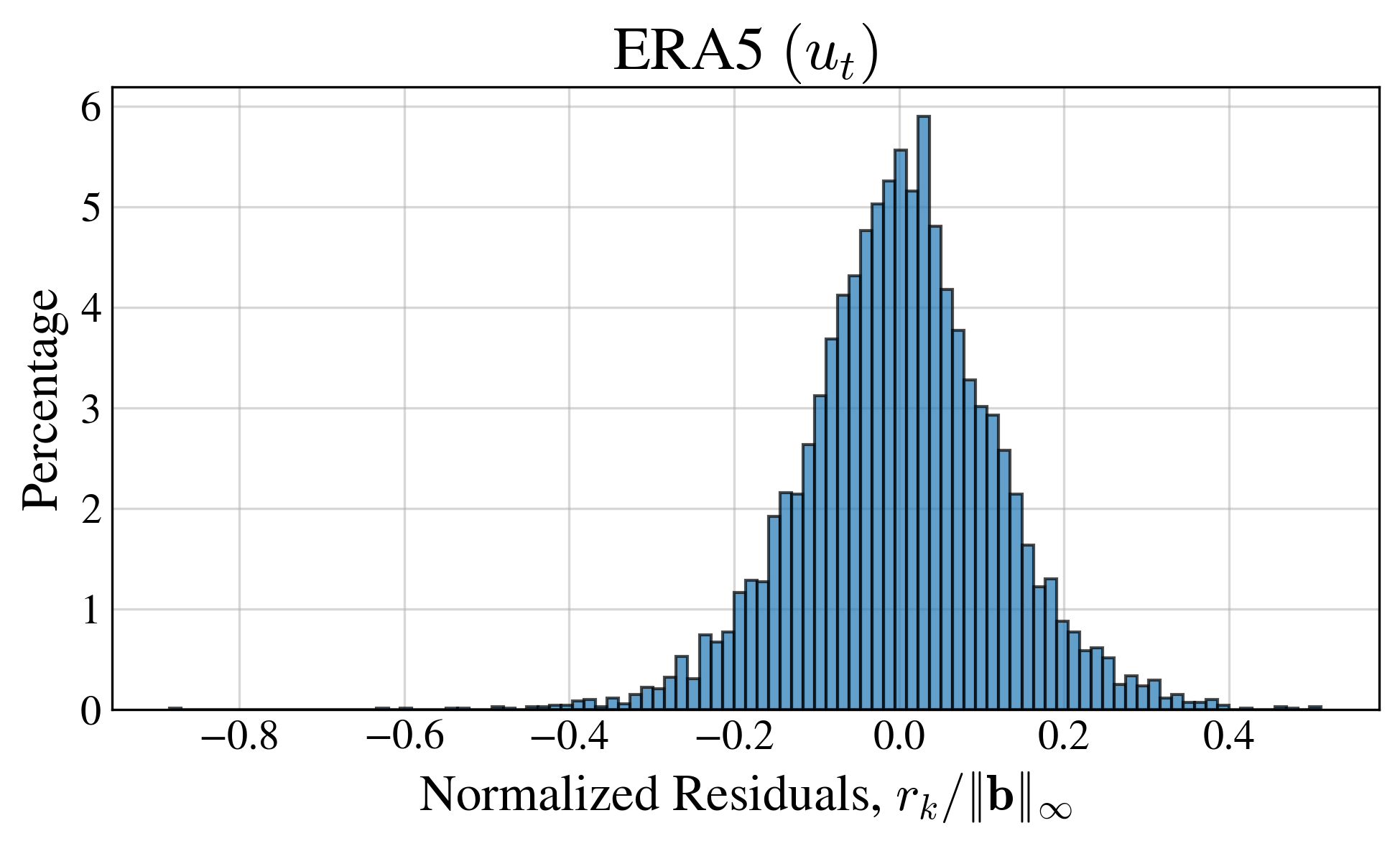}
    \includegraphics[width=0.485\linewidth]{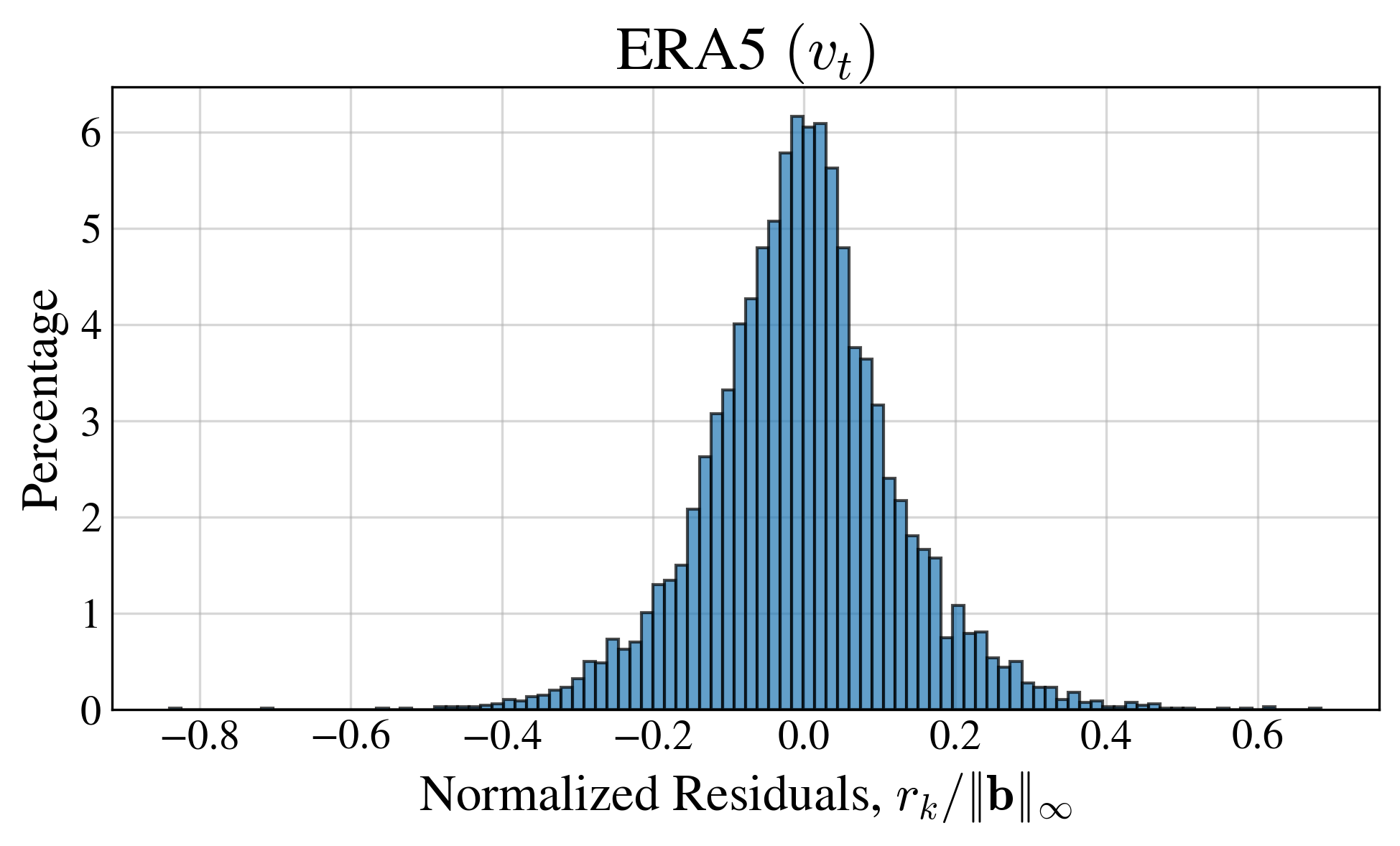}
    \includegraphics[width=0.495\linewidth]{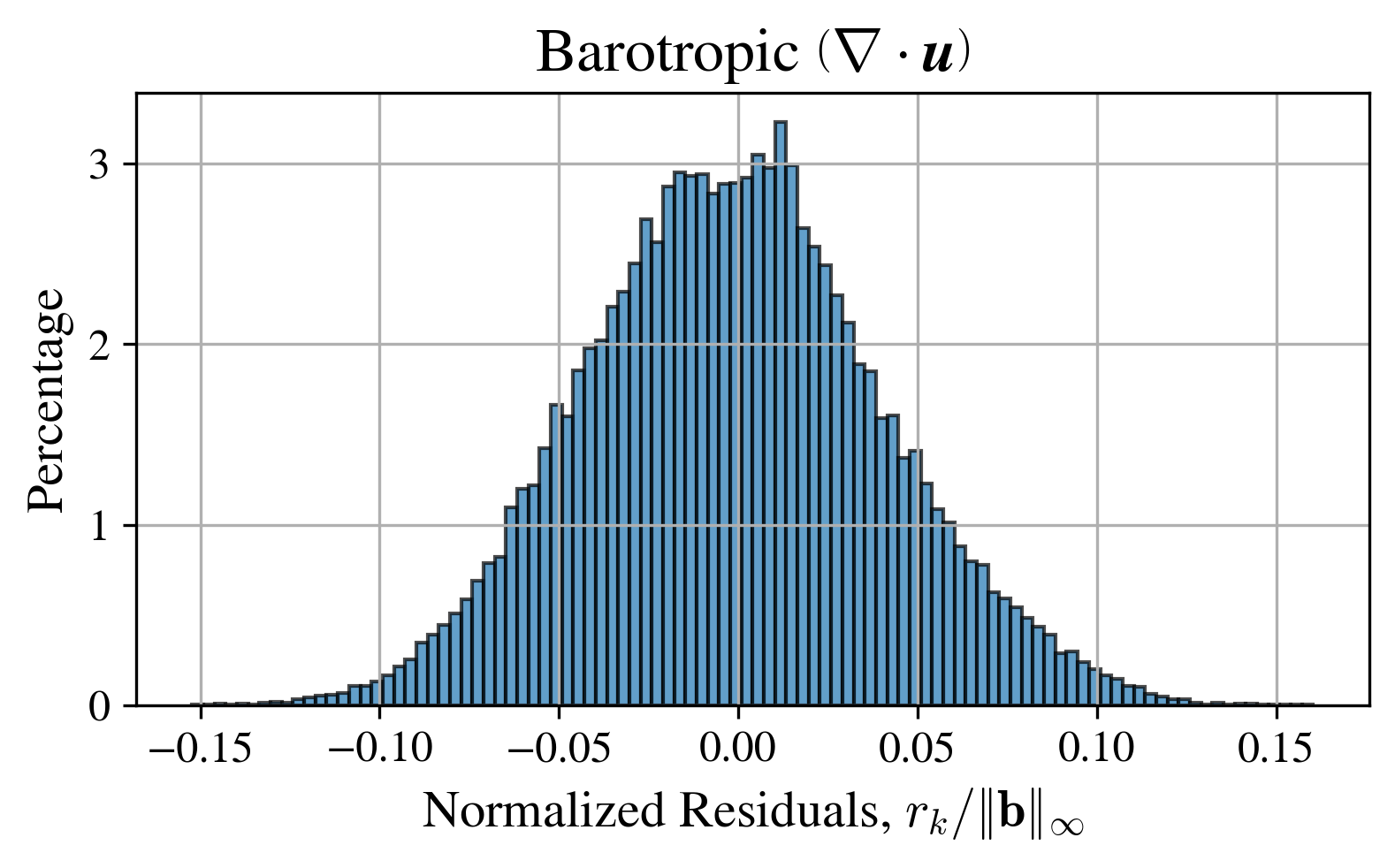}
    \caption{{\color{red}Additional residual plots, complementing the results shown in Figure~\ref{fig:numerical_results_1}.}}
    \label{fig:additional_residuals}
\end{figure}

\begin{table}[htb]
\begin{center}
\begin{tabular}{||c c c c||} 
\hline
\textbf{Symbol} & \textbf{Definition} & \textbf{Type of Object} &  \textbf{Lives In}\\ [0.5ex] 
\hline\hline
$n$ & Spatial dimension & Constant & $n \in \mathbb{N}$ \\ [1ex]
$\mathcal{X}$ & Spatial domain & Compact set & $\mathcal{X} \subset \mathbb{R}^n$ \\ [1ex]
$T$ & Final time (observations) & Constant & $T > 0$ \\ [1ex]
$\tau$ & Final time (training data) & Constant & $0 < \tau < T
$ \\ [1ex]
\hline
$\boldsymbol{x}$ & Position, $(x_1, \dots, x_n)$ & Vector coord. & $\boldsymbol{x} \in \mathcal{X}$ \\ [1ex]
$t$ & Time & Scalar coord. & $t \in [0,T]$ \\ [1ex]
$\varphi$ & Longitude & Scalar coord. & $\varphi \in [0, 2\pi)$ \\ [1ex]
$\theta$ & Latitude & Scalar coord. & $\theta \in \big[\!-\frac{\pi}{2}, \frac{\pi}{2}\big]$ \\ [1ex]
$r$ & Altitude & Scalar coord. & $r \geq 0$ \\ [1ex]
$\eta$ & Hybrid vertical coord. & Scalar coord. & $\eta \geq 0$ \\ [1ex]
\hline
$u$ & Zonal wind & Scalar field & $u(\boldsymbol{x},t) \in \mathbb{R}$ \\ [1ex]
$v$ & Meridional wind & Scalar field & $v(\boldsymbol{x},t) \in \mathbb{R}$ \\ [1ex]
$w$ & Vertical wind & Scalar field & $w(\boldsymbol{x},t) \in \mathbb{R}$ \\ [1ex]
$\boldsymbol{u}$ & Horizontal wind vel., $(u,v)$ & Vector field & $\boldsymbol{u}(\boldsymbol{x},t) \in \mathbb{R}^2$ \\ [1ex]
$\boldsymbol{v}$ & Full wind velocity, $(u,v,w)$ & Vector field & $\boldsymbol{v}(\boldsymbol{x},t) \in \mathbb{R}^3$ \\ [1ex]
\hline
$\Phi$ & Geopotential & Scalar field & $\Phi(\boldsymbol{x},t) \in \mathbb{R}$ \\ [1ex]
$\zeta$ & Relative vorticity & Scalar field & $\zeta(\boldsymbol{x},t) \in \mathbb{R}$ \\ [1ex]
$\omega$ & Potential vorticity & Scalar field & $\omega(\boldsymbol{x},t) \in \mathbb{R}$ \\ [1ex]
$\Omega$ & Planetary angular vel. rate & Physical const. & - \\ [1ex]
$f$ & Coriolis parameter, $2\Omega\sin(\theta)$ & Scalar field & $f(\theta) \in \mathbb{R}$ \\ [1ex]
$\boldsymbol{f}$ & Coriolis force, $(-fv, fu)$ & Vector field & $\boldsymbol{f}(\theta) \in \mathbb{R}^n$ \\ [1ex]
\hline
\end{tabular}
\end{center}
\caption{\color{red} Reference table for symbols used to refer to empirical quantities.}
\label{table:notation_physical}
\end{table}

\begin{table}[htb]
\begin{center}
\begin{tabular}{||c c c c||} 
\hline
\textbf{Symbol} & \textbf{Definition} & \textbf{Type of Object} &  \textbf{Lives In}\\ [0.5ex] 
\hline\hline
$d$ & Number of state variables & Constant & $d \in \mathbb{N}$ \\ [1ex]
$M$ & Number of observations & Constant & $M \in \mathbb{N}$ \\ [1ex]
$I$ & Number of candidate operators & Constant & $I \in \mathbb{N}$ \\ [1ex]
$J$ & Number of candidate functions & Constant & $J \in \mathbb{N}$ \\ [1ex]
$K$ & Number of query points & Constant & $K \in \mathbb{N}$ \\ [1ex]
$\mathcal{X}_{\Delta}\!\!\times\!T_{\Delta}$ & Discrete domain, $\{(\boldsymbol{x}_m, t_m)\}_{m=1}^{M}$ & Discrete set & - \\ [1ex]
\hline
$\boldsymbol{\ell}$ & Support radii, $(\ell_{x_1}, \dots, \ell_t)$ & Tuple & $\boldsymbol{\ell} \in \mathbb{N}^{n+1}$ \\ [1ex]
$p_i$ & Test function degree & Constant & $p_i \in \mathbb{N}$ \\ [1ex]
$\phi_i$ & Test function component & Scalar field & $\phi_i \in C_c^{p_i}(\mathbb{R})$ \\ [1ex]
$\psi$ & Test function, $\phi_t(t)\prod_{i=1}^{n}\phi_i(x_i)$ & Scalar field & $\psi \in C_c^p(\mathbb{R}^{n+1})$ \\ [1ex]
\hline
$u_l$ & State variable & Scalar field & $u_l \in L^1\big(\mathbb{R}^{n+1}\big)$ \\ [1ex]
$\boldsymbol{u}$ & State vector, $[u_1, \dots, u_d]$ & Vector field & $\boldsymbol{u} \in L^1\big(\mathbb{R}^{n+1}\big)$ \\ [1ex]
$\mathbf{u}_l$ & Discretized variable, vec$\{u_l(\boldsymbol{x}_m, t_m)\}$ & Vector & $\mathbf{u}_l \in \mathbb{R}^M$ \\ [1ex]
$\mathbf{U}$ & Discretized state vector, $[\mathbf{u}_1, \dots, \mathbf{u}_d]$ & Matrix & $\mathbf{U} \in \mathbb{R}^{M \times d}$ \\ [1ex]
$\mathcal{U}$ & Set of observations, $\{\boldsymbol{u}(\boldsymbol{x}_m, t_m)\}_{m=1}^{M}$ & Discrete set & - \\ [1ex]
\hline
$\mathbf{\Theta}$ & Library of candidate terms & Matrix & $\mathbf{\Theta} \in \mathbb{R}^{M \times IJ}$ \\ [1ex]
$\mathbf{w}$ & Coefficients, $[\mathbf{w}_1, \dots, \mathbf{w}_d]$ & Matrix & $\mathbf{w} \in \mathbb{R}^{IJ \times d}$ \\ [1ex]
$\mathbf{b}$ & $b_{kl} = (\psi_t * u_l)(\boldsymbol{x}_k, t_k)$ & Matrix & $\mathbf{b} \in \mathbb{R}^{K \times d}$ \\ [1ex]
$\mathbf{G}$ & $G(i,j)_k = \big(\mathcal{D}^i\psi_t * f_j(\boldsymbol{u})\big)(\boldsymbol{x}_k, t_k)$ & Matrix & $\mathbf{G} \in \mathbb{R}^{K \times IJ}$ \\ [1ex]
$\mathbf{r}$ & WSINDy residual, $\mathbf{b} - \mathbf{G}\mathbf{w}^{\star}$ & Matrix & $\mathbf{r} \in \mathbb{R}^{K \times d}$ \\ [1ex]
\hline
$\mathcal{L}$ & Regularized loss function & Scalar field & $\mathcal{L}(\mathbf{w}) \geq 0$ \\ [1ex]
$E_{\infty}$ & $\max_j |w^{\star}_j - w_j^{\rm{true}}| / |w_j^{\rm{true}}|$ & Constant & $E_{\infty} \geq 0$ \\ [1ex]
TPR & True positive ratio & Constant & $0 \leq {\rm{TPR}} \leq 1$ \\ [1ex]
$\overline{\mathcal{E}}_F$ & Avg. relative error at $t=T$ & Constant & $\overline{\mathcal{E}}_F \geq 0$ \\ [1ex]
$t_{\rm{tol}}$ & $\min\big\{t : \overline{\mathcal{E}}(t) \geq 0.1\big\}$ & Constant & $t_{\rm{tol}} \geq 0$ \\ [1ex]
\hline
\end{tabular}
\end{center}
\caption{\color{red} Reference table for symbols used to refer to numerical or WSINDy-related quantities.}
\label{table:notation_numerical}
\end{table}

\end{document}